\documentclass[11pt]{article}

\usepackage[utf8]{inputenc}
\usepackage[T1]{fontenc}
\usepackage{lmodern}
\usepackage{amsmath,amssymb}
\usepackage{booktabs}
\usepackage{geometry}
\usepackage{hyperref}
\usepackage{xcolor}
\usepackage{enumitem}
\usepackage[ruled,vlined,linesnumbered]{algorithm2e}
\usepackage{microtype}
\usepackage{caption}
\usepackage{tikz}
\usetikzlibrary{arrows.meta,positioning,fit,backgrounds,calc}
\usepackage{pgfplots}
\pgfplotsset{compat=1.18}
\usepackage{placeins}
\usepackage{graphicx}
\hypersetup{colorlinks=true, linkcolor=blue!50!black, citecolor=blue!50!black,
            urlcolor=blue!50!black}

\newcommand{\fs}{\texttt{fast\_send}}
\newcommand{\snd}{\texttt{send}}

\title{\textbf{Adapting the Actor Model of Concurrency for\\
High-Frequency Trading}\\
\large Synchronous Message Delivery (\texttt{fast\_send}) and a Tick-to-Book Latency Study}

\author{%
  Vincent Maciejewski\\
  \small M2 Technologies\\
  \small \texttt{mayeski@gmail.com}
}

\date{September 2026}

\begin{document}
\maketitle

\begin{abstract}
The many benefits of the actor model --- per-actor state isolation, freedom
from data races, and sequential single-message reasoning --- have been
recognized for decades, yet the model has
long been dismissed as unsuitable for high-frequency trading (HFT). 
The reasoning was that actors imply many
threads, many mailbox queues, and a message --- with its heap allocation, queue
operations, scheduler dispatch, and context switch --- for every interaction:
overhead incompatible with a microsecond latency budget. This paper argues that
the dismissal is wrong for co-located actors, and supports the argument not only
analytically but with a deployed, measured implementation --- \emph{kaspar-hft},
an open-source C++20 framework that adapts the actor model into a form suitable
for HFT and is its reference implementation. The extensions are four: \fs{}, a
synchronous delivery
mechanism in which the sending thread runs the receiver's handler inline and
takes the reply as a value; actor \emph{groups}, which co-schedule a set of actors
on one thread behind a single shared mailbox; selectable mailbox queue
implementations matched per actor to its write pattern; and a memory pool that
removes per-message allocation overhead.

The central mechanism is \fs{}, whose defining property is \emph{receiver
transparency}: the handler is written identically for synchronous and
asynchronous delivery and does not depend on which was used, which thread
executes it, or whether the sender is blocked. Synchronous delivery
across a chain of grouped actors runs entirely on the first actor's thread,
reducing scheduler context switches from $O(N)$ on the asynchronous path to
$O(1)$; a thread-local call-chain test detects cyclic synchronous invocation on
one thread before any lock is taken, while a cycle spread across threads is
outside what it can see and deadlocks; and five delivery operations expose the
latency/robustness spectrum behind one transparent interface. Microbenchmarks put
the synchronous round trip at tens of nanoseconds --- roughly two orders of
magnitude below the ungrouped asynchronous path --- and a single \fs{} dispatch at
about the cost of one unpredictable virtual call (about $5\times$ a direct call).

We then measure the framework in a real HFT setting: socket-to-book
(``tick-to-book'') latency on a live CME market-data feed for ES, NQ, and ZN
futures, measured with software timestamps from socket read to book update. When
the socket-reader thread decodes each packet and updates the book itself,
book-update medians are $0.8$--$1.1~\mu$s, and a $\sim\!9$~ns \fs{} hop is about
$1\%$ of that. When the reader instead hands each datagram to a second thread
through an asynchronous mailbox, the median rises to $\sim\!7~\mu$s, and the
tail grows with in-packet position and, on rare bursts, mailbox-queue backlog,
driven by the market's non-Poisson, strongly clustered arrival process, which we
characterize in a companion paper. Finally, the shared-queue group yields a
production/simulation duality: the same actor code runs unchanged in live trading
and in deterministic backtest. A saturated-throughput and contention evaluation
is left to future work.
\end{abstract}

\vspace{0.5em}

% ---------------------------------------------------------------
\section{Introduction}
% ---------------------------------------------------------------

In the actor model \cite{hewitt1973,agha1986}, a program is a collection of
isolated actors, each with private state, a message handler, and a mailbox.
Actors communicate only by asynchronous message passing: a sender places a
message on the receiver's mailbox and proceeds without waiting, and the receiver
dequeues and processes messages one at a time on its own thread. This discipline
is what gives the model its guarantees --- isolation, sequential per-actor
processing, and location transparency --- and it is essential when actors are
distributed across machines.

When the communicating actors reside in the same process on the same machine,
however, the asynchronous path carries overhead that the isolation guarantee
does not require. A single request-and-reply between two co-located, in-process actors
incurs, on conventional implementations:

\begin{itemize}[leftmargin=1.4em]
  \item \textbf{Heap allocation.} Because the message must outlive the sender's
  current continuation and persist until processed at an indeterminate later
  time, it must be heap-allocated, with the associated allocation cost,
  fragmentation, and garbage-collection pressure.
  \item \textbf{Queue operations.} Each message is enqueued by the sender and
  dequeued by the receiver, each requiring synchronization on the mailbox.
  \item \textbf{Scheduling and context switches.} The receiver must be scheduled
  onto a thread to run, incurring a context switch; a context switch typically
  costs on the order of $10^3$--$10^4$ processor cycles and pollutes cache and
  translation lookaside buffer (TLB) state \cite{li2007,mcvoy1996,bendersky2018}.
  \item \textbf{Round-trip latency.} A reply repeats the entire path in reverse,
  so a request-and-reply crosses the mailbox twice and the scheduler twice.
\end{itemize}

For a same-process pair, all of this is avoidable in principle: the sender's own
thread could run the receiver's handler directly and take the reply as a return
value. The difficulty is doing so without breaking the abstraction the actor
model exists to provide. A direct handler invocation --- calling the receiver's
handler as an ordinary method --- discards the receiver's exclusive-execution
guarantee and exposes the receiver as a passive object.
The design problem is to obtain the
performance of a direct, same-thread call while preserving, from the receiver's
point of view, the semantics of message passing.

The central mechanism this paper introduces is \fs{}, a synchronous delivery
mechanism that resolves that
tension. The sending thread acquires exclusive access to the receiving actor and
executes the receiver's handler directly, on the sender's thread, returning the
reply as a value. The receiver's handler is identical for synchronous and
asynchronous delivery and does not depend on which was used --- the property we
call \emph{receiver transparency}. Because synchronous invocations can form
cycles, and a cycle deadlocks a non-reentrant lock or overflows the stack under
a reentrant one, the mechanism carries a lightweight cyclic-invocation detector
that fires before any lock is taken.\footnote{kaspar-hft does not implement this
detector yet: \fs{} takes the receiver's mutex directly, and its only check is a
debug-build assertion against an actor synchronously sending to itself, so a
synchronous cycle currently deadlocks.} It sees cycles on one thread; a cycle
spread across threads needs a different mechanism (Section~\ref{sec:cycle}).
\fs{} is implemented in \emph{kaspar-hft}, an
open-source C++20 actor framework for high-frequency
trading;\footnote{\url{https://github.com/vincent212/kaspar-hft}} all code and
measurements in this paper are drawn from that repository.

\paragraph{What was changed to make a conventional actor framework HFT-ready.}
This is a paper about a working trading system before it is a paper about
concurrency: the question it answers is whether a live tick-to-book path can be
built on actors without paying a latency penalty for the discipline. Making a
textbook actor framework fit that path took four concrete changes, each removing
one source of the per-message cost that the mainstream one-thread-per-actor,
one-mailbox-per-actor design pays. \emph{First}, \fs{}, the synchronous
delivery mechanism above, which runs a co-located receiver's handler inline on
the sender's thread and eliminates the allocation, queue operations, scheduling,
and context switch of an asynchronous send. \emph{Second}, actor \emph{groups},
which co-schedule a set of actors on one thread behind a single shared mailbox,
so an entire servicing stage collapses onto one core with no cross-thread hop.
\emph{Third}, per-actor selection of the mailbox queue implementation, matched to
each actor's write pattern (single bursty producer, high fan-in, or grouped),
because no one queue is fastest in every regime. \emph{Fourth}, a memory pool
that removes per-message allocation and, more importantly, the multi-millisecond
allocator tail that would otherwise land directly in the latency distribution.
The remainder of the paper develops each in turn and then measures, on a live CME
feed: with decode and book update on one thread, socket-to-book medians are about
$1~\mu$s, of which a \fs{} hop is about one percent.

\paragraph{Contributions.}
\begin{enumerate}[leftmargin=1.6em]
  \item A synchronous, same-thread actor-delivery mechanism (\fs{}) that
  eliminates the heap allocation, queue operations, scheduling, and context
  switches of the asynchronous path for co-located actors, and returns the reply
  as a value (Section~\ref{sec:mech}).
  \item A precise statement of \emph{receiver transparency}: the handler
  executes identical code in both modes and does not depend on the delivery
  mode, the executing thread, or whether the sender is blocked
  (Section~\ref{sec:transparency}). It is a property of how handlers are
  written, not one the framework or the language enforces.
  \item A thread-local call-chain membership test that detects cyclic
  synchronous invocation on one thread before lock acquisition, together with a
  refined (actor, message-class) variant that admits callback patterns while
  still catching unbounded recursion (Section~\ref{sec:cycle}).
  \item A family of five delivery operations spanning asynchronous through
  spin-wait synchronous to best-effort-with-async-fallback, unified behind one
  transparent handler interface (Section~\ref{sec:family}).
  \item Actor \emph{groups}, which co-schedule a set of actors on one thread
  behind one shared mailbox so that an intra-group send never crosses a core; the
  single shared queue additionally yields a production/simulation duality, in
  which the same actor code runs unchanged in live trading and in deterministic
  backtest (Sections~\ref{sec:groups},~\ref{sec:duality}).
  \item Per-actor selection of the mailbox queue implementation over four
  multi-producer/single-consumer designs, with a selection rule derived from
  measurement across the grouped, bursty, and high-fan-in regimes
  (Sections~\ref{sec:queues-mech},~\ref{sec:qtypes}).
  \item A memory pool that removes per-message heap allocation and, more
  importantly, the allocator's multi-millisecond tail from the latency
  distribution (Section~\ref{sec:pool}).
  \item An evaluation of the mechanism in isolation: an analytical account of the
  per-message work eliminated and microbenchmarks of the C++ implementation,
  including drain-side batching ($1.63\times$) and memory pooling that together
  raise asynchronous throughput $2.46\times$, and the pool's
  two-order-of-magnitude reduction of the latency-tail maximum
  (Section~\ref{sec:perf}).
  \item A live socket-to-book case study on CME market data
  (Section~\ref{sec:casestudy}): with decode and book update on the
  socket-reader thread, book-update medians are $0.8$--$1.1~\mu$s, about
  $100\times$ a \fs{} hop; with a cross-thread hand-off between reader and
  decoder, latency decomposes into a floor plus an in-packet serialization slope,
  and the study separates mailbox-queue occupancy from in-packet position and relates the tail to a non-Poisson, strongly
  clustered arrival process. A saturated-throughput evaluation is left to
  future work.
\end{enumerate}

% ---------------------------------------------------------------
\section{Origins and history of the actor model}
\label{sec:history}
% ---------------------------------------------------------------

The actor model was introduced by Hewitt, Bishop, and Steiger in 1973
\cite{hewitt1973} as a single uniform notion --- the \emph{actor}, an entity that
communicates only by sending messages to other actors --- and was given a rigorous
semantics over the following decade by Greif \cite{greif1975}, Clinger
\cite{clinger1981}, and, most influentially, Agha \cite{agha1986}, who recast the
model on the primitives \emph{create}, \emph{send}, and \emph{become}. A
distinguishing feature of these semantics is \emph{unbounded nondeterminism}: a
message is guaranteed to arrive eventually but with no bound on how long it may be
delayed, so the set of possible outcomes can be infinite (in contrast to
\emph{bounded} nondeterminism, where a fixed process count and prompt scheduling
keep that set finite). The most influential early \emph{deployed} system with
actor-like structure was Erlang \cite{armstrong1996}, whose designers arrived at
similar principles largely independently --- Armstrong later placed Erlang closer
to communicating sequential processes than to the actor model \cite{armstrong2007}.
The subsequent decades produced a broad ecosystem of actor runtimes --- on the
JVM, .NET, C++, Rust, Swift, and elsewhere --- which we survey next.

% ---------------------------------------------------------------
\section{A survey of actor-model implementations}
\label{sec:survey}
% ---------------------------------------------------------------

This section surveys the major actor runtimes along the three axes most relevant
to \fs{}: how actors are mapped to operating-system threads, whether and how a
synchronous or request--reply path is offered, and what is known of their
performance. The consistent findings across the ecosystem --- that actors are
multiplexed M:N over a small worker pool, and that no mainstream system runs the
\emph{receiver's} handler on the \emph{sender's} thread --- are what locate the
contribution of \fs{}. Table~\ref{tab:survey} summarizes the comparison; the
per-system detail and performance figures follow. The later related-work
discussion (Section~\ref{sec:related}) revisits the systems closest to \fs{} to
sharpen the distinction; here the account is descriptive.

\begin{table}[t]
\centering
\caption{Actor-to-thread mapping and synchronous messaging across
implementations. The final column --- whether the sender's thread ever executes
the receiver's message handler --- is negative for every surveyed system and
positive only for \fs{}.}
\label{tab:survey}
\footnotesize
\setlength{\tabcolsep}{4pt}
\begin{tabular}{p{2.2cm} p{1.8cm} p{3.1cm} p{3.0cm} c}
\toprule
System & Runtime & Actor $\to$ thread mapping & Synchronous / request--reply & Sender runs handler? \\
\midrule
Erlang/OTP, Elixir & BEAM VM & M:N green processes; one scheduler thread per core; per-scheduler run queues, work stealing & \texttt{gen\_server:call} blocks caller, returns a value; handler runs in the server process & No \\
Akka & JVM & thread-pool dispatchers over a fork--join pool; actors not 1:1 with threads & \texttt{ask} (\texttt{?}) returns a \texttt{Future}; \texttt{tell} (\texttt{!}) is fire-and-forget & No \\
Akka.NET, Orleans & .NET & thread-pool / turn-based scheduling; grains single-threaded per activation & grain methods return \texttt{Task<T>}; always asynchronous & No \\
CAF & C++ & cooperative work-stealing scheduler over $P{=}\#\text{cores}$ workers; \texttt{detached} actors get a thread & \texttt{request} returns a response handle; \texttt{.then} continuation, non-blocking & No \\
SObjectizer & C++ & agents bound to dispatchers supplying worker threads/pools & async by default; \texttt{so5extra} \texttt{request\_reply} blocks; runs on provider's dispatcher & No \\
Actix & Rust / Tokio & actors are tasks on Arbiter event-loop threads (M:N) & \texttt{send} returns a \texttt{Request} future; \texttt{do\_send} fire-and-forget & No \\
Pony & Pony RT & work-stealing scheduler, one thread per core; per-actor heaps & none --- behaviours are asynchronous by design & No \\
Swift actors & Swift RT & serial executor per actor over a cooperative pool & cross-actor access requires \texttt{await}; synchronous cross-actor calls are prohibited & No \\
Ray, Dapr, Proto.Actor & distributed & one worker/process or mailbox per actor; M:N pools & remote call returns a future; turn-based dispatch & No \\
\midrule
\fs{} (this work) & C++/Rust & per-actor exclusive-access lock; sender's thread runs the handler & blocking call returns a value on the sender's thread & \textbf{Yes} \\
\bottomrule
\end{tabular}
\end{table}

\paragraph{BEAM: Erlang and Elixir.} The Erlang runtime multiplexes very large
numbers of lightweight processes --- each with a private heap and mailbox ---
onto a small set of scheduler threads, one per logical core by default, with a
per-scheduler run queue and work-stealing migration \cite{beambook}. Scheduling
is preemptive by a reduction count: a process runs until it consumes its budget
(4000 reductions by default) and is then descheduled, which bounds tail latency
\cite{beambook}. The synchronous path, \texttt{gen\_server:call}, blocks the
caller for up to a timeout and returns a value, but the request is placed in the
\emph{server} process's mailbox and handled by \texttt{handle\_call/3} on the
server's own scheduler; the caller merely performs a selective receive on the
reply. The caller's thread never runs the callee's handler
\cite{armstrong1996}.

\paragraph{JVM: Akka.} Akka does not bind actors to threads. A dispatcher binds
a set of actors to a fork--join thread pool, and an actor occupies a thread only
while it has messages to process \cite{akka}. A \texttt{throughput} setting caps
how many messages an actor processes before the thread moves to the next actor;
this mailbox batching is the principal latency/throughput lever. Akka's own
microbenchmark reports on the order of 50 million messages per second on a single
machine in a ping-pong configuration, a figure the authors show is dominated by
that batching setting; it is a vendor microbenchmark, not an application
measurement, and should be read as such \cite{letitcrash}. The request--reply
operator \texttt{ask} returns a \texttt{Future}/\texttt{CompletionStage} rather
than a value and never blocks the caller on the handler; Akka's documentation
explicitly discourages blocking inside actors because it starves the shared pool
\cite{akka}.

\paragraph{.NET: Orleans and Akka.NET.} Orleans introduced the ``virtual actor''
(grain) abstraction, in which the runtime activates grains on demand and
schedules them turn-based and single-threaded over a shared pool; every grain
method is asynchronous and returns a \texttt{Task}, so a caller obtains a promise
and the grain body runs on the grain's own activation context, never on the
caller's thread \cite{orleans}. The Orleans team reported the Halo~4 presence
service sustaining roughly 130{,}000 requests per second across 25 servers with
near-linear scaling to 125 servers \cite{orleansmsr2014}. A recent independent
empirical comparison of Akka, Akka.NET, and Orleans on commodity cloud hardware
found Akka.NET and Orleans comparable and both ahead of Akka in a
non-distributed setup, with Orleans strongest in the distributed setup
\cite{camilleri2023}.

\paragraph{C++: CAF and SObjectizer.} The C++ Actor Framework schedules
cooperatively-run actors over a work-stealing pool of one worker per core, with
a lock-free mailbox; actors that must block opt into a \texttt{detached} mode
that gives them a dedicated thread \cite{caf2014}. Its request--reply operation,
\texttt{request}, does not occupy a CAF thread while waiting: the requesting
actor simply processes no other message until the response arrives, and installs
a typed continuation via \texttt{.then}; the caller does not execute the
callee's handler \cite{caf2014}. On the Savina cross-language benchmarks
\cite{imam2014savina}, a
recent study found CAF only marginally ahead of Akka on average (about
$1.09\times$ at 20 threads) and behind it on nine of sixteen benchmarks
\cite{linguafranca2023}, so the common claim that native C++ actors dominate the
JVM is more nuanced than it is often stated. SObjectizer likewise binds agents to
dispatchers backed by worker-thread pools and is asynchronous by default; its
companion library offers a synchronous \texttt{request\_reply}, but the handler
still runs on the service provider's dispatcher, not the caller's thread.

\paragraph{Rust: Actix.} Actix \cite{actix} runs actors as tasks on Arbiter event loops atop
the Tokio runtime, multiplexing many actors per thread. \texttt{send} returns a
\texttt{Request} future that resolves to the handler's result, and
\texttt{do\_send} is fire-and-forget; in both cases the handler runs on the
target actor's Arbiter, not the sender's. (The widely-cited Actix TechEmpower
web-framework results concern \texttt{actix-web}, not the actor library, and are
not evidence about actor-message performance.)

\paragraph{Pony.} Pony is the clearest case of the mainstream design point.
Behaviours are asynchronous by construction --- there is no synchronous send at
all --- so request--reply must be built by hand from callback messages. Pony
runs a work-stealing scheduler with one thread per core and gives each actor its
own heap, which enables a fully concurrent, per-actor garbage collector (ORCA)
that needs neither read nor write barriers, relying instead on reference
capabilities for data-race freedom and on causal message delivery for
correctness \cite{ponyorca2017}. Its performance comes from cheap actors,
local GC, and never blocking.

\paragraph{Swift.} Swift's language-level actors (SE-0306) give each actor a
serial executor drawn from a cooperative thread pool. Cross-actor access must be
\texttt{await}-ed and is therefore asynchronous; the compiler \emph{prohibits}
synchronous cross-actor calls outright \cite{swift2021}. Swift actors are
reentrant: a suspended actor may interleave other work before resuming.

\paragraph{Distributed runtimes.} Ray, Dapr, and Proto.Actor extend the actor
idea across machines. A Ray actor is by default a single worker executing
methods one at a time, and a call returns an object reference (a future); Dapr
enforces turn-based, single-threaded access to each actor via a per-actor lock
and dispatches method calls over HTTP/gRPC; Proto.Actor processes each mailbox
one message at a time on a shared thread-pool dispatcher. In all three the
invocation is asynchronous and the sender never runs the receiver's handler.

\paragraph{Two structural facts.} Two observations hold across the entire
survey. First, on \emph{thread mapping}: the dominant design multiplexes many
actors M:N over a small pool of worker or scheduler threads, because assigning a
dedicated operating-system thread to each actor does not scale to large actor
populations; the few one-thread-per-actor cases (CAF's \texttt{detached} actors,
Ray's threaded actors, process-per-actor deployments) are reserved for blocking
or isolation needs and are understood as exceptions. Second, on
\emph{synchronous messaging}: where a synchronous or request--reply path exists
at all, it is emulated over asynchronous mailbox delivery --- the caller either
blocks in a receive (Erlang) or, more commonly, receives a future/promise (Akka,
Orleans, CAF, Actix, Ray) --- and in every case the receiver's handler runs on
the receiver's own scheduled thread or context. Some systems (Pony, Swift's
cross-actor calls) forbid a synchronous call altogether. Among the runtimes
surveyed here we did not find one that executes the receiver's handler on the
sender's thread under the receiver's exclusive-access lock. That combination --- cross-thread synchronous
execution under receiver transparency --- is the design point \fs{}
occupies. It needs protection against synchronous cycles, which
Section~\ref{sec:cycle} provides for cycles on one thread.

% ---------------------------------------------------------------
\section{Why actors? The design case and its cost}
\label{sec:benefits}
% ---------------------------------------------------------------

Before addressing performance, one question must be settled: why adopt the
actor model at all, when a program can already spawn operating-system threads
and share memory under mutual exclusion, or drive concurrency through an event
loop such as \texttt{boost::asio}? Those mechanisms are not deficient in
capability. The case for actors is not that they compute something the
alternatives cannot; it is what the discipline \emph{removes} --- the classes
of error it prevents by construction.

\paragraph{What the discipline removes.} Because each actor owns its state and
communicates only by messages, there is no mutable state shared between actors,
and therefore no memory-level data race, the most common serious
failure of lock-based concurrency. Empirical study of the field supports this:
data races and deadlocks are described as ``two mistakes that are hard to make
with actors'' \cite{tasharofi2013}, and a comprehensive study of 186 real-world
actor bugs confirms that actor programs are markedly less susceptible to
low-level data races and deadlocks than multithreaded code, even as they admit
their own higher-level bug classes \cite{bagherzadeh2020,torreslopez2018}. The
programmer reasons about one message at a time against a consistent private
state, sequentially, and never writes a lock. The synchronization apparatus that
dominates the correctness burden of hand-rolled concurrent code --- choosing
lock granularity, establishing a global lock order to avoid deadlock, placing
memory barriers, and reasoning about every interleaving that a thread scheduler
or an \texttt{asio} callback chain admits --- simply does not appear in an actor
program. Fault isolation follows from the same isolation: a failing actor can be
restarted by a supervisor without corrupting the state of others, the
foundation of the ``let it crash'' reliability discipline \cite{armstrong2003}.

These properties belong to the asynchronous model. The synchronous delivery of
Section~\ref{sec:mech} gives one of them up: a \fs{} holds the receiver's lock
while its handler runs, and a chain of \fs{} calls holds several actors' locks at
once. A cycle of synchronous calls on one thread is caught by the detector of
Section~\ref{sec:cycle}; a cycle across threads, where thread~1 holds $A$ and
waits for $B$ while thread~2 holds $B$ and waits for $A$, deadlocks, exactly as a
lock-ordering error does in lock-based code. The asynchronous path keeps the
guarantee; the synchronous path trades part of it for latency.

\paragraph{The cost.} These properties are not free. The message-passing
discipline that provides isolation is
also the source of the per-message overhead this paper addresses:
heap allocation, mailbox queue operations, scheduling, and context switches
(Section~\ref{sec:perf}). This overhead is the cost of isolation, and on
latency-critical paths it has historically led engineers to abandon the actor
model and return to shared memory and locks. The conflict between the model's
safety and its per-message cost is the subject of this paper.

\paragraph{Determinism and simulation.} A further benefit follows from the same
isolation and is especially important in a trading system. Because an actor's behavior
is a function of the message sequence on its mailbox and nothing else --- no
shared state, and no wall-clock or global-random side channel once the actor
obtains time and randomness as messages --- an actor is deterministically
reproducible: the same input sequence yields the same states, outputs, and faults
on every run. This makes the actor a natural unit of \emph{simulation}. When
actors are co-scheduled behind one queue (Section~\ref{sec:micro}), that single
queue is fed by live sources in production or replayed from a recorded or
synthetic sequence in simulation, and the actors cannot tell the two apart. The
same code is therefore both a real-time execution engine and a deterministic
simulator: a trading strategy is developed and backtested against recorded market
data, debugged by replaying the exact sequence that produced a fault, and
deployed to production without modification. Shared-memory and callback-based
designs do not offer this without effort --- their behavior depends on thread
interleavings and ambient clocks that a simulator must otherwise mock away,
whereas for actors the property follows from the message discipline. The duality,
and its use in developing an execution algorithm, are the subject of
Section~\ref{sec:duality}.

\paragraph{The generation argument.} A further benefit has become relevant only
recently. An actor
is a small, regular unit of code: a class with private state and one handler per
message type, whose entire observable contract is the set of messages it accepts
and the replies it returns. This regularity makes actors unusually amenable to
automatic construction by large language model (LLM) code generators, which are
fluent in the paradigm: a model can scaffold an actor, write its message
handlers, and --- because an actor's behavior is a function from an incoming
message to a reply with no hidden shared state --- generate self-contained unit
tests that drive messages in and assert on the replies, without the elaborate
harness that testing shared-memory code requires. The contrast with the
alternative is documented. Generating \emph{correct} lock-based
concurrent code is hard even for strong models: a dedicated benchmark for
concurrent code generation finds that current LLMs struggle to produce correct
synchronization and repeatedly introduce concurrency errors
\cite{concur2026}. The actor discipline eliminates the hazard that
both human authors and code generators handle worst --- shared mutable state
guarded by hand-written locks --- so actor code plausibly is not only faster to
produce but more robust and closer to correct than a hand-rolled alternative
built on a custom locking scheme. We state this as a design hypothesis, not a
result: we do not evaluate generated-code quality here. The supporting evidence
is indirect --- the demonstrated difficulty of the lock-based alternative
\cite{concur2026} together with the measured data-race freedom of the actor
approach \cite{tasharofi2013,bagherzadeh2020} --- and a systematic evaluation of
LLM-generated actor implementations, for correctness, concurrency safety, and
performance, is left to future work.

\paragraph{Removing the co-located cost.} Taken together, the
benefits of the actor model --- data-race freedom, sequential per-actor
reasoning, fault isolation, and amenability to automatic generation
and testing --- have always been offset by a performance penalty that put the
model impractical on latency-critical paths. If that penalty is removed for
the common case of co-located actors, the trade-off changes: the isolation, the
absence of hand-written locking, and the amenability to code generation are
retained, while the per-message cost approaches that of a direct call. Providing
that path, safely and without breaking the abstraction, is the purpose of \fs{},
described next.

% ---------------------------------------------------------------
\section{Background and related work}
\label{sec:related}
% ---------------------------------------------------------------

The actor model \cite{hewitt1973,agha1986} specifies asynchronous message
passing as the sole means of inter-actor communication, and foundational and
subsequent treatments present this asynchrony as essential to isolation, fault
tolerance, and location transparency \cite{agha1986,sutter2005}. Several systems
provide a second, synchronous-looking communication path; each differs from \fs{}
in a way that bears on receiver transparency or on multi-threaded execution.

\paragraph{Single-threaded synchronous calls.} The E language
\cite{miller2005} offers an immediate call (\texttt{.}) returning a value and an
eventual send (\texttt{<-}) returning a promise. E vats are single-threaded, so
the immediate call is a conventional method invocation with no lock, no
cross-thread execution, and no possibility of deadlock; the model does not
extend to actors running on multiple threads. AmbientTalk \cite{ambienttalk2007}
similarly builds on communicating event loops in which cross-actor communication
is always asynchronous. Actor4j \cite{actor4j2018} optimizes messaging between
actors on the \emph{same} thread, and reliable-actor systems such as the
Kubernetes Application Runtime (KAR)
\cite{kar2021} bypass the queue only for an actor's calls to itself, not for
cross-actor calls.

\paragraph{Blocking that still uses the receiver's thread.} Erlang's
\texttt{gen\_server:call} \cite{armstrong1996} blocks the caller, but the
request still traverses the receiver's mailbox and is processed on the
receiver's own process; the caller's thread never executes the receiver's
handler. Akka's ask operator \cite{akka} returns a future and remains
asynchronous internally, routing through the mailbox, and the framework
explicitly discourages blocking inside actors. Orleans \cite{orleans} presents
grain methods in call syntax but is asynchronous underneath
(\texttt{Task}-returning), with no caller-thread execution of the grain. Swift's
actors \cite{swift2021} require \texttt{await} for cross-actor access and
prohibit synchronous cross-actor calls at the compiler level. 
The C++ Actor Framework (CAF) \cite{caf2014} dispatches handlers on a scheduler's worker pool, so the
caller thread does not run the target's handler.

\paragraph{Direct execution without message-passing semantics.} The Active
Object pattern \cite{activeobject1996} interposes a proxy, activation queue, and
scheduler so that the client's thread never runs the servant's method on a
separate thread. The Monitor Object pattern \cite{posa2} does run a method on the
caller's thread under a lock --- sharing the caller-thread-execution mechanism
with \fs{} --- but a monitor is a passive object invoked by a direct method
call: it has no mailbox, offers no asynchronous alternative, and does not present
the callee as an isolated actor that cannot observe how it was reached.

\paragraph{Deterministic reactor concurrency.} A separate recent line of work
constrains, rather than accelerates, actor-style concurrency: Lingua Franca
\cite{linguafranca2023} schedules reactors under a logical-time model to make
concurrent execution deterministic. That objective is orthogonal to \fs{}, which
addresses the per-message cost of co-located delivery and preserves receiver
transparency; the two could be combined, since \fs{} does not alter the logical
ordering of message handling.

\subsection{Positioning}
\label{sec:positioning}

Across these systems, a synchronous path either (i) is confined to a single
thread or same-thread actors (E, AmbientTalk, Actor4j), (ii) blocks the caller
while still executing on the receiver's thread through the mailbox (Erlang,
Akka, Orleans, CAF), (iii) is prohibited outright (Swift), or (iv) abandons
message-passing semantics for direct method calls on a passive object (Active
Object, Monitor Object). What distinguishes \fs{} is the combination of
(a) cross-thread synchronous execution --- the sender's thread runs the
receiver's handler while holding the receiver's exclusive-access lock ---
(b) under receiver transparency, with (c) a cyclic-invocation detector that
catches synchronous cycles on one thread before they deadlock or overflow the
stack; cycles across threads are outside what it can see
(Section~\ref{sec:cycle}). The remainder of the paper develops these three
elements.

% ---------------------------------------------------------------
\section{Requirements of an HFT data path}
\label{sec:requirements}
% ---------------------------------------------------------------

Before adapting the actor model, we state what an HFT data path requires of any
concurrency substrate it is built on. Two requirements dominate, and together
they set the design target for the rest of the paper.

\paragraph{Performance, specifically the tail.} For an HFT data path performance
is the overriding requirement, and the figure of merit is not the mean latency
but the tail --- the $99$th, $99.9$th, and $99.99$th percentiles. Market data
decoded late and orders sent late are worthless or produce a loss, so a strategy
is judged on its worst-case reaction time, not its typical one. A concurrency
layer that is fast on average but occasionally slow is therefore unacceptable in
a way that has no analogue in throughput-oriented systems.

\paragraph{Arrivals are clustered, so tails are intrinsic.} Market data does not
arrive smoothly. Market events arrive in bursts rather than evenly --- the feed is
strongly non-Poisson, as ours confirms (Section~\ref{sec:arrival}) --- and the
exchange coalesces the events available at each send into a single datagram, so a
burst of events becomes a long packet. A long packet converts into a
long latency tail, because its messages are decoded in sequence. A path designed
for the \emph{average} arrival rate would therefore be mis-designed; the system
must stay fast enough under a burst that its servicing queue does not build,
which makes the quantities to engineer the per-message service time and its tail,
not the mean throughput.

\paragraph{Consequence for the concurrency layer.} The two requirements combine
into a single target for the messaging substrate: it must add negligible
per-message cost \emph{and no latency tail of its own}, while preserving the
actor model's isolation and single-message reasoning
(Section~\ref{sec:benefits}). The four changes of the next section are built to
meet this target, and the case study of Section~\ref{sec:casestudy} then confirms
the per-hop cost is small: with decode and book update on one thread the
book-update median is about $1~\mu$s, of which a \fs{} hop is about $1\%$. It
also shows the other side of the same argument: when each datagram instead
crosses a thread through an asynchronous mailbox, the median is about $7~\mu$s and part of the tail is mailbox-queue
backlog, so where the path crosses threads is a design decision in its own right.

% ---------------------------------------------------------------
\section{Adapting the actor model for HFT}
\label{sec:extensions}
% ---------------------------------------------------------------

The survey of Section~\ref{sec:survey} fixed the mainstream design point:
actor runtimes deliver every message asynchronously, give each actor its own
mailbox, and multiplex many actors M:N over a small pool of worker threads. On
that design even a request--reply between two actors in the same process pays a
heap allocation, an enqueue and a dequeue, a scheduler dispatch, and a context
switch --- and, for the reply, all of it again. That is the correct design
\emph{across machines}, where the asynchrony buys isolation, fault tolerance,
and location transparency (Section~\ref{sec:benefits}); it is over-provisioned
for actors that share an address space, where the sender's own thread could run
the receiver's handler directly.

To meet the requirements of Section~\ref{sec:requirements}, we make four changes
to that design. Each is a departure from one standard-actor assumption, and each
removes a specific per-message cost --- or its tail --- while preserving the
isolation, sequential per-actor reasoning, and the receiver transparency that
motivate the model (Table~\ref{tab:contrast}). The first and central change, \fs{},
occupies Sections~\ref{sec:mech}--\ref{sec:family}; the remaining three ---
actor groups, selectable mailbox queues, and a memory pool --- follow in
Sections~\ref{sec:groups}--\ref{sec:pool}.

\begin{table}[h]
\centering
\caption{The four changes as departures from the mainstream actor design. Each
removes a co-located per-message cost; none alters the isolation or the
message-passing semantics a handler observes.}
\label{tab:contrast}
\small
\begin{tabular}{p{2.5cm} p{4.7cm} p{5.0cm}}
\toprule
Dimension & Mainstream actor model & This work (kaspar-hft) \\
\midrule
Delivery & asynchronous only; the handler runs later on the receiver's scheduled thread & synchronous inline delivery on the sender's thread (\fs{}) for co-located actors; asynchronous retained for remote \\
Actor $\to$ thread & one mailbox per actor, M:N over a worker pool, one scheduled turn per message & a \emph{group} co-schedules actors on one thread behind one shared mailbox \\
Mailbox queue & one fixed queue implementation & four implementations, selectable per actor to its write pattern \\
Message memory & heap-allocated per message & stack-allocated for synchronous requests; pooled when it must outlive the call \\
\bottomrule
\end{tabular}
\end{table}

\subsection{Synchronous delivery: \fs{}}
\label{sec:mech}

Each actor has a message handler, a mailbox, private state, and a synchronization
mechanism that guarantees at most one thread executes its handler at a time
(a mutex by default; a lock-free primitive is admissible). Two delivery
operations share this structure.

\paragraph{Asynchronous delivery (\snd{}).} The sender heap-allocates the
message, enqueues it on the receiver's mailbox under the mailbox's
synchronization, and returns immediately with no value. Later, a thread assigned
to the receiver dequeues the message, acquires the receiver's exclusive-access
lock, and runs the handler. A reply, if any, is heap-allocated, wrapped in a
response message, and enqueued on the sender's mailbox.

\paragraph{Synchronous delivery (\fs{}).} The sender may allocate the message on
its \emph{stack}, because the sending thread blocks until processing completes
and the message's lifetime is therefore bounded by the call. The sender's thread
acquires the \emph{receiver's} exclusive-access lock (blocking if another thread
holds it), pushes the receiver onto a thread-local call chain
(Section~\ref{sec:cycle}), and invokes the receiver's handler directly on the
sender's thread. The handler runs its identical code and optionally calls the
reply operation; on return, the receiver is popped from the call chain, the lock
is released, and \fs{} returns the reply value (or null if none was set). A
stack-allocated message is reclaimed when the sender's frame unwinds.

In both modes the receiver's exclusive-access lock is held for the duration of
handler execution, so the ``one message at a time'' guarantee holds identically
under the conservative detection policy; the permissive (actor, message-class)
refinement of Section~\ref{sec:cycle} admits same-actor re-entrancy and relaxes it.
The distinction from a naive direct call is that the synchronous path holds the
receiver's exclusive-access lock and routes the reply through the same operation
as the asynchronous path; it is delivery of a message that happens to execute on
the sender's thread.

\paragraph{Priority semantics.} A \fs{} invocation does \emph{not} join the
receiver's mailbox behind messages already queued there. It executes as soon as it
acquires the receiver's exclusive-access lock, so synchronous delivery has
\emph{priority} over queued asynchronous delivery: if the receiver's mailbox holds
$m_1,m_2,m_3$ and a sender issues $\fs(m_4)$, then once the lock is free $m_4$'s
handler runs first, and $m_1,m_2,m_3$ run afterward on the receiver's own turn.
This is intentional. \fs{} is a high-priority, synchronous request/response for a
caller that needs the receiver's current state without waiting for the mailbox to
drain; it is not a faster spelling of an ordinary asynchronous \snd{}, and it
therefore reorders that one invocation ahead of the queue. The exclusive-access
lock is still respected: if the receiver is mid-handler when $\fs(m_4)$ arrives,
the sender blocks on the lock and $m_4$ runs after that handler returns but still
ahead of the remaining queue. A caller that requires strict FIFO with respect to
other senders must use \snd{}.

\paragraph{Unified reply routing.} A single reply operation serves both modes and
selects its behavior from delivery-context information rather than from any
argument the handler supplies. Under asynchronous delivery it heap-allocates the
value, wraps it in a response message, and enqueues it on the sender's mailbox;
under synchronous delivery it heap-allocates the value and stores it in a
return-value location that the blocked \fs{} reads and returns. The value is
heap-allocated in \emph{both} modes deliberately: if the synchronous path could
stack-allocate the reply while the asynchronous path could not, the handler would
have to know the mode, defeating transparency. The reply may be invoked at most
once per message.\footnote{kaspar-hft does not enforce this yet: a second reply
to a synchronous delivery replaces the first, and a second reply to an
asynchronous one sends a second message.} An opt-in optimized reply stores the
value without heap allocation for code paths known to be
synchronous-only;\footnote{kaspar-hft does not implement the optimized reply
yet.} it raises an error if the message was delivered asynchronously, and by
design breaks transparency, so its use is explicit.

\paragraph{Chained synchronous execution.} If actor $A$ synchronously invokes
$B$, which synchronously invokes $C$, and so on, the entire chain executes on
$A$'s thread, which holds the exclusive-access locks of all actors in the chain
simultaneously and, when those locks are uncontended, performs no context
switches. Such an $N$-actor synchronous chain incurs $O(1)$ context switches,
against $O(N)$ for the equivalent asynchronous chain, and executes on a single
core, avoiding the cross-core cache-line migration of state that the asynchronous
path incurs. A contended lock in the chain blocks its thread, which reintroduces
a context switch at that step.

\subsection{Receiver transparency}
\label{sec:transparency}

Receiver transparency is the property that the receiving actor's handler is
written identically for both delivery modes and does not depend on (i) whether it
was reached synchronously or asynchronously, (ii) which thread is executing it,
or (iii) whether the sender is blocked awaiting a reply. It is a property of how
handlers are written, not isolation that the framework or the language enforces. A
handler that looks can infer the mode: C++ cannot stop it from inspecting its own
execution context --- its thread identifier, the address of a local variable (and
so which thread's stack it is on), or the time --- and some message fields the
framework sets are filled differently in the two modes (element 4). Transparency
means that a handler has no reason to look, and that its code does not change
between modes. It is what keeps \fs{} within the actor
abstraction rather than reducing it to a locked method call.

Four elements bear on it:

\begin{enumerate}[leftmargin=1.6em]
  \item \textbf{Identical interface.} The handler receives a message and invokes
  the same reply operation in both modes; there is no mode flag, and no
  conditional branch in the handler depends on the delivery mode.
  \item \textbf{Exclusive access in both modes.} The receiver's exclusive-access
  lock is held throughout handler execution whether the handler runs on the
  receiver's own thread (asynchronous) or on the sender's thread
  (synchronous), so the observable execution environment --- single-writer,
  sequential --- is the same.
  \item \textbf{Mode-independent reply and allocation.} The unified reply routes
  by delivery context with no handler involvement, and the reply value is
  heap-allocated in both modes, so neither the reply's behavior nor allocation
  side effects reveal the mode.
  \item \textbf{No delivery flag for handlers.} The flag recording how a message
  was delivered is private to the framework. Other message fields are not
  mode-neutral: the queue depth and the ``mailbox empty'' flag describe the
  receiver's mailbox on asynchronous delivery and have fixed values on synchronous
  delivery, and the destination field is set only on asynchronous delivery. They
  exist for instrumentation and routing; a handler whose logic depends on them
  gives up transparency.
\end{enumerate}

Two consequences follow. First, the delivery mode becomes a \emph{sender-side}
optimization: a sender may upgrade an asynchronous send to a synchronous one with
no change, recompilation, or redeployment of the receiver. Second, the mechanism
is distinct from a direct procedure call, in which the callee is a passive object
invoked on the caller's stack with no mailbox and no exclusive-execution
guarantee; here the receiver remains an isolated actor that perceives the arrival
of a message and is unaware that the sender's thread is the one running it.

\subsection{Cyclic-invocation detection}
\label{sec:cycle}

\paragraph{What the detector cannot see: cycles across threads.} The call chain below is
per thread, so it detects only a cycle that stays on one thread. It cannot see a
cycle spread across threads: thread~1 runs a \fs{} to $A$ whose handler does a
\fs{} to $B$, while thread~2 runs a \fs{} to $B$ whose handler does a \fs{} to
$A$. Each thread's check passes --- $B \notin \{A\}$ and $A \notin \{B\}$ ---
and each thread then blocks on the lock the other holds. \fs{} is callable from
any thread, so this case is real. Catching it needs shared state rather than a
thread-local chain: a global wait-for graph checked before blocking, a global
order on actor locks, or a bounded try-lock that falls back to asynchronous
delivery (the \texttt{fast\_send\_void} policy of Section~\ref{sec:family}).
Choosing among these is future work (Section~\ref{sec:future}).

Synchronous invocation introduces a failure mode that asynchronous delivery does
not have. If synchronous calls form a cycle --- $A \to B \to C \to A$ --- then
under a non-reentrant lock the thread waits to acquire an actor's lock it already
holds, and the system deadlocks; under a reentrant lock the handler re-enters
with the lock's recursion count incremented, producing unbounded nested
execution that overflows the stack and exposes inconsistent intermediate state.
Neither is acceptable, and the cycle cannot be discovered after the offending
lock is taken --- by then the thread is already stuck.

\subsubsection{Call-chain membership test}

The mechanism maintains, per thread, a \emph{call chain}\footnote{kaspar-hft
does not implement the call chain yet.}: the ordered set of
actors whose handlers are currently executing on that thread via synchronous
invocation. Because it is thread-local, it requires no synchronization and resets
automatically between tasks. It is initialized when a thread begins processing a
dequeued (asynchronous) message --- the dequeued actor is its first entry --- and
each synchronous invocation pushes its target on entry and pops it on completion.
The chain may be represented as a stack or list, a hash set for $O(1)$ membership
queries, a bitmap indexed by actor identifier, or a hybrid.

Before acquiring the target's lock, a synchronous send queries whether the target
is already present in the chain. If it is not, the send proceeds normally; if it
is, a cyclic invocation has been detected, and the mechanism signals it --- by
raising a cyclic-invocation exception carrying the target identity and the full
chain, or by returning an error indicator --- \emph{without} acquiring the lock
or invoking the handler. Algorithm~\ref{alg:cycle} states the check. Because the
query precedes lock acquisition and reads only thread-local state, its overhead
is negligible against the cost of a lock acquisition and a handler invocation,
and the chain depth is small in practice.

\begin{algorithm}[t]
\caption{Synchronous send with cyclic-invocation detection}
\label{alg:cycle}
\DontPrintSemicolon
\SetKwInOut{State}{thread-local}
\State{call chain $C$ (set of actors executing synchronously on this thread)}
\BlankLine
\KwIn{target actor $B$, message $m$}
\BlankLine
\If{$B \in C$}{
  \Return cyclic-invocation signal (exception or error indicator)
     \tcp*{before any lock is taken}
}
acquire $B$'s exclusive-access lock\;
$C \gets C \cup \{B\}$\;
$r \gets B.\text{handler}(m)$ \tcp*{runs on this (the sender's) thread}
$C \gets C \setminus \{B\}$\;
release $B$'s exclusive-access lock\;
\Return $r$\;
\end{algorithm}

\subsubsection{Refined detection with message classes}

Membership keyed on actor identity alone is conservative: it rejects legitimate
callbacks in which $A$ synchronously invokes $B$ and $B$ synchronously invokes a
\emph{different} handler of $A$. A refined mode\footnote{kaspar-hft does not
implement the refined mode yet.} keys the call chain on
(actor, message-class) pairs, where the message-class discriminator may be a type
tag, a class identifier, a handler function pointer, or any code-path
discriminator. A cycle is signaled only when the same pair recurs. This admits a
callback pattern such as $A/\texttt{compute} \to B/\texttt{validate} \to
A/\texttt{validated}$ --- three distinct pairs --- while still catching genuine
unbounded recursion such as $A/\texttt{foo} \to B/\texttt{bar} \to
A/\texttt{foo}$, in which a pair repeats. Because permissive keying allows an
actor's state to be re-entered by a nested handler, reentrancy safety in that
mode is the programmer's responsibility. The policy --- conservative
(actor-only) or permissive (pair-based) --- is configurable system-wide,
per-actor, or per-invocation.

\subsection{A family of delivery operations}
\label{sec:family}

The same transparent handler interface supports five delivery
operations\footnote{kaspar-hft implements \snd{} and \fs{}; it does not
implement the other three, or the adaptive variant, yet.} that
differ only in how the sender behaves when the receiver's lock is contended or a
cycle is detected (Table~\ref{tab:five}). All synchronous variants perform the
cyclic-invocation check of Section~\ref{sec:cycle} before attempting the lock.

\begin{table}[h]
\centering
\caption{The five delivery operations. All share one receiver-transparent
handler interface; they differ in sender-side behavior under contention and
cycles.}
\label{tab:five}
\small
\begin{tabular}{lccll}
\toprule
Operation & Blocks & Returns & Lock unavailable & Cycle detected \\
\midrule
\snd{} (async)          & no  & ---   & (enqueues)            & --- \\
\fs{} (blocking sync)   & yes & value & waits (sleep on lock) & exception \\
\texttt{fast\_send\_spin} & yes & value & waits (busy-wait)     & exception \\
\texttt{fast\_send\_x}    & no  & value & \texttt{ActorBusy} exception & exception \\
\texttt{fast\_send\_void} & no  & ---   & async fallback        & async fallback \\
\bottomrule
\end{tabular}
\end{table}

\begin{itemize}[leftmargin=1.4em]
  \item \textbf{\fs{}} blocks on the receiver's lock by sleeping, and raises a
  cyclic-invocation exception on a cycle. It is the default synchronous
  operation.
  \item \textbf{\texttt{fast\_send\_spin}} is identical except that it
  busy-waits on the lock, optionally issuing a CPU \texttt{pause} instruction between
  attempts, trading CPU cycles for lower wake latency by avoiding the scheduler.
  It suits short critical sections on latency-sensitive paths with spare cores.
  \item \textbf{\texttt{fast\_send\_x}} attempts the lock without blocking:
  on success it runs the handler and returns the value; on contention it raises
  an \texttt{ActorBusy} exception, distinct from the cyclic-invocation
  exception, so the caller can fall back, retry, or propagate. Its execution
  time is bounded.
  \item \textbf{\texttt{fast\_send\_void}} attempts the lock without blocking and
  returns no value: on success it runs the handler; on contention \emph{or} on a
  detected cycle it enqueues the message asynchronously and returns, copying a
  stack-allocated message to the heap first if necessary. Because it never blocks
  on the target lock, the operation itself cannot be the waiting party in a
  lock-wait deadlock; the handler it runs on success may still block or recurse.
  It fits notification, forwarding, pipeline, and broadcast patterns. The ratio
  of synchronous completions to asynchronous fallbacks is a useful contention
  metric.
  \item An adaptive variant spins for a bounded count and then transitions to
  blocking acquisition, combining low latency under light contention with
  bounded CPU use under heavy contention.
\end{itemize}

Selection among the operations is a sender-side decision governed by whether the
caller needs a return value, can tolerate an unbounded wait, and has idle
processor capacity available; the receiver's handler is unaffected in every case.

\subsection{Actor groups}
\label{sec:groups}

A \emph{group} co-schedules a set of actors on a single thread behind one shared
mailbox: the members do not own separate queues, and an asynchronous \snd{}
between two actors in the same group never crosses a core --- it is a push and a
pop on a queue owned by the running thread, with no cross-thread wakeup. This is
the standard low-latency actor optimization (same-thread messaging in Actor4j, a
shared dispatcher in Akka; Section~\ref{sec:survey}), and it is the substrate
\fs{} builds on: grouping the actors of a servicing stage onto one thread lets an
$N$-actor pipeline collapse to a single inline \fs{} call chain
(Section~\ref{sec:hftarch}). The single shared queue has a second consequence
beyond latency --- it imposes one total order on every message entering the
group, which is what lets the same actor code run unchanged in live trading and
in deterministic backtest (Section~\ref{sec:duality}).
Section~\ref{sec:micro} reports the grouped round-trip cost.

\subsection{Selectable mailbox queues}
\label{sec:queues-mech}

An actor's mailbox is not a fixed data structure. It is a \texttt{std::variant}
over four multi-producer/single-consumer queue implementations, chosen per actor
to match its write pattern: \texttt{BQueue} (a mutex and condition variable
around a ring buffer), \texttt{BQueueBatched} (the same lock, but the consumer
drains the whole queue under one acquisition), \texttt{ShardedBQueue} ($N$
independently locked lanes joined by an atomic cursor), and \texttt{LockFreeMPSC}
(a bounded lock-free Vyukov ring, in which a producer reserves a slot with one
atomic operation). Because the variant holds the concrete type, \texttt{push} and
the drain loop are devirtualized and inlined. No one implementation is fastest in
every regime, so the choice is exposed rather than fixed;
Section~\ref{sec:qtypes} measures the four and gives the selection rule.

\subsection{Memory pool}
\label{sec:pool}

The message-passing discipline heap-allocates a message per interaction, and on a
hot path that allocation is both a per-message cost and, worse, a source of tail
latency when the general allocator takes its occasional slow path. Two mechanisms
remove it. A synchronous \fs{} request needs no allocation at all: the sender
blocks until the handler returns, so the message can live on the sender's stack
(Section~\ref{sec:mech}). A message that must outlive the call is drawn from a
fixed-size \texttt{MemoryPool} rather than \texttt{new}/\texttt{delete}, which
lowers the mean cost and, more importantly, removes the allocator's
multi-millisecond tail. Section~\ref{sec:micro} measures both effects.

% ---------------------------------------------------------------
\section{A representative HFT system architecture with \fs{}}
\label{sec:hftarch}
% ---------------------------------------------------------------

This section shows where the four changes sit in a real market-data-to-order
pipeline, and how the servicing chain is kept fast enough to hold its queue near
empty under the clustered arrivals of Section~\ref{sec:requirements}; the case
study of Section~\ref{sec:casestudy} then measures what the framework contributes
on a live feed.

\begin{figure}[t]
\centering
\resizebox{\linewidth}{!}{%
\begin{tikzpicture}[
    font=\small, >=Latex,
    box/.style={draw, rounded corners, minimum height=8mm, minimum width=15mm, align=center, fill=blue!5},
    reader/.style={box, fill=green!10},
    outbox/.style={box, fill=orange!12},
    lbl/.style={font=\scriptsize},
    seclbl/.style={font=\scriptsize\itshape, text=black!70},
]
\node[reader] (sra) {Socket\\reader A};
\node[reader] (srb) [below=7mm of sra] {Socket\\reader B};
\coordinate (mid) at ($(sra)!0.5!(srb)$);
\node[box] (buf)  [right=18mm of mid] {Buffer\\(queue)};
\node[box] (dec)  [right=15mm of buf] {Arbitrate\\+ decode};
\node[box] (book) [right=15mm of dec] {Order\\book};
\node[box] (sig)  [right=15mm of book] {Signal};
\node[outbox] (om)   [right=16mm of sig] {Order\\handler};

% incoming feeds
\draw[->] ($(sra)+(-14mm,0)$) -- node[lbl,above]{A} (sra);
\draw[->] ($(srb)+(-14mm,0)$) -- node[lbl,below]{B} (srb);
% async sends into the buffer (context switch)
\draw[->] (sra) -- node[lbl,sloped,above]{send} (buf);
\draw[->] (srb) -- node[lbl,sloped,below]{send} (buf);
% fast_send chain (single thread)
\draw[->] (buf)  -- node[lbl,above]{\texttt{fast\_send}} (dec);
\draw[->] (dec)  -- node[lbl,above]{\texttt{fast\_send}} (book);
\draw[->] (book) -- node[lbl,above]{\texttt{fast\_send}} (sig);
% async send out (context switch)
\draw[->] (sig)  -- node[lbl,sloped,above]{send} (om);
\draw[->] (om) -- node[lbl,above]{TCP} ($(om)+(15mm,0)$);

\begin{scope}[on background layer]
\node[draw,dashed,rounded corners,fit=(sra)(srb),inner sep=3mm] (s1) {};
\node[draw,dashed,rounded corners,fit=(buf)(dec)(book)(sig),inner sep=3mm] (s2) {};
\node[draw,dashed,rounded corners,fit=(om),inner sep=3mm] (s3) {};
\end{scope}
\node[seclbl,below=1mm of s1.south] {Capture: 2 threads};
\node[seclbl,below=1mm of s2.south] {Decode / book / signal: 1 thread};
\node[seclbl,below=1mm of s3.south] {Order out: 1 thread};
\end{tikzpicture}%
}
\caption{Thread architecture of a representative HFT pipeline. The only
cross-thread hops on the critical
path are the two asynchronous \snd{}s (reader$\to$buffer and
signal$\to$order-handler); each such hop is the expensive operation, measured at
$\sim\!3.4\,\mu$s per round trip (Table~\ref{tab:transport}), because it wakes
another thread. The entire decode/book/signal servicing chain instead collapses
onto a single thread via \fs{}, each hop costing $\sim\!9$~ns --- about one
mispredicted virtual call (Table~\ref{tab:dispatch}), three orders of magnitude less than a thread-crossing \snd{} (see
Table~\ref{tab:transport}) --- so it incurs no queue and no context switch. The
per-message service time of that servicing chain, and hence how fast its queue
drains, is set by the aggregate cost of these \fs{}s. The case study
(Section~\ref{sec:casestudy}) measures the capture-to-book part of this path
both with the reader decoding and updating the book on its own thread
(book-update medians $0.8$--$1.1~\mu$s) and with the reader$\to$buffer
hand-off shown here (median $\sim\!7~\mu$s).}
\label{fig:hftarch}
\end{figure}
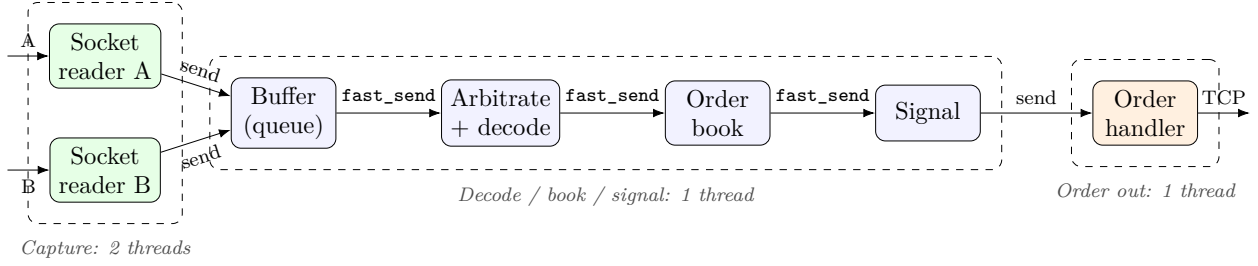

Figure~\ref{fig:hftarch} shows the thread architecture of a representative HFT
market-data-to-order pipeline, drawn from the kaspar-hft reference
implementation. CME's MDP~3.0 market data arrives as two redundant multicast
feeds, channel~A and channel~B. The system decomposes into three sections
connected by exactly two cross-thread hops.

\paragraph{Section 1 --- capture (two threads).} One socket-reader actor per
channel does nothing but drain its UDP socket and hand each datagram to the next
stage by an \emph{asynchronous} \snd{}. It must be asynchronous: a reader that
made a synchronous, blocking call would stop draining its socket while the callee
ran and would drop packets under a burst --- defeating the reader's only purpose.
In kaspar-hft the reader pre-allocates message buffers and batches whatever the
socket has available before sending, and its handoff to the buffer stage is an
async \snd{}, never a \fs{}.

\paragraph{Section 2 --- decode, book, and signal (one thread).} The datagrams
land in a buffer actor --- the queue --- whose service handler forwards each
message down the remainder of the chain by \fs{}: arbitration between the A and B
feeds, Simple Binary Encoding decode, order-book construction, and signal
generation all run synchronously, in sequence, on this one thread, with no
intervening queue and no context switch between stages. This is the chained
synchronous execution of Section~\ref{sec:mech}: an $N$-stage chain with
$O(1)$ rather than $O(N)$ context switches. This is the queue-servicing code
whose aggregate per-message time determines how fast the market-data queue
drains.

\paragraph{Section 3 --- order out (one thread).} When the signal decides to
act, it hands an order to an order-management actor by an asynchronous \snd{};
that actor performs the TCP/iLink write. This hop is asynchronous by design: the
network write takes several microseconds, and blocking the decode-and-book thread
on it would stall servicing of the market-data queue and let it back up --- the
exact failure the architecture exists to avoid.

\paragraph{Two cross-thread handoffs.} On the critical path from wire to order
there are just two cross-thread hops --- reader$\to$buffer and
signal$\to$order-out. Each is an asynchronous \snd{} that may wake another thread,
and when it does the wakeup and any attendant context switch cost a few
microseconds --- a thread-to-thread switch measures near $1.2$--$1.5~\mu$s on a
pinned core, rising to a few microseconds once cache-working-set effects are
included \cite{li2007,bendersky2018,mcvoy1996}. Everything between them is
collapsed onto a single thread by
\fs{}. The goal is not to remove those two handoffs; they are cheap and
necessary, because the capture and transmit stages must not block the servicing
stage. The goal is to make the servicing stage between them fast enough that its
queue never grows, which is what \fs{} does by collapsing the chain onto one
thread. The mechanism's contribution is thus best measured not as a per-call
latency but as the aggregate per-message service time it sustains on the
servicing chain.

% ---------------------------------------------------------------
\section{Performance analysis}
\label{sec:perf}
% ---------------------------------------------------------------

The synchronous path removes, rather than optimizes, the per-message operations
of the asynchronous path. Table~\ref{tab:elim} summarizes the operations
eliminated for a co-located delivery.

\begin{table}[h]
\centering
\caption{Per-message operations on the asynchronous path and their status on the
synchronous path, for co-located actors. This is an analytical account of work
removed.}
\label{tab:elim}
\small
\begin{tabular}{lcc}
\toprule
Per-message operation & Asynchronous & Synchronous \\
\midrule
Message heap allocations           & 1--3 & 0 request (stack-alloc.), 1 reply \\
Mailbox queue operations           & 2 (enqueue $+$ dequeue) & 0 \\
Scheduler invocations              & 1--2 & 0 \\
Context switches (request/reply)   & 2--4 & 0 \\
Context switches, $N$-actor chain  & $O(N)$ & $O(1)$ \\
\bottomrule
\end{tabular}
\end{table}

Each eliminated item has a concrete cost. A context switch is on the order of
$10^3$--$10^4$ processor cycles and evicts cache and TLB state
\cite{li2007,mcvoy1996,bendersky2018}. Heap allocation
of short-lived messages produces fragmentation and, in managed runtimes,
collection pauses; stack allocation reclaims the message on frame unwind with no
allocator or collector involvement. Removing the mailbox operations removes their
synchronization traffic on the memory bus; quantifying that reduction requires
measurement and is left to future work. The chained-execution
result --- $O(1)$ rather than $O(N)$ context switches across an $N$-actor
synchronous chain --- is the largest structural effect for deep call graphs.

\paragraph{Context: measured cost of the asynchronous path.} To indicate the
magnitude of the overhead the synchronous path targets, we measured the
asynchronous messaging path of the same actor runtime
at $5.5$ to $13.6$ million round trips per second after
successive batching and memory-pooling optimizations; in the saturated
throughput regime a message costs on the order of $74$~ns, while in the sparse
regime an individual send costs several microseconds ($4$--$7~\mu$s), with
condition-variable wakeups costing $1$--$3~\mu$s each. These are measurements of
the \emph{asynchronous} path and bound the per-message costs that synchronous
delivery removes; they are not measurements of \fs{}.

\paragraph{Batching the asynchronous path.} The $2.46\times$ batching-and-pooling
gain above is worth
a note, because it is the cross-thread counterpart of \fs{}: where \fs{} removes
the mailbox from the co-located same-thread path, batching amortizes it on the
cross-thread path that must stay asynchronous (the two capture and transmit hops
of Figure~\ref{fig:hftarch}). The baseline pays a mailbox-lock acquisition on
every push and pop, an actor-lock acquisition per dispatch, and a
condition-variable wakeup per send. Draining the whole queued burst under a
\emph{single} mailbox-lock acquisition, and signaling the condition variable only
on the empty-to-non-empty transition, raised throughput from $5.5$ to $8.96$
million round trips per second ($1.63\times$); adding the memory pool reached
$13.6$ million ($\sim\!74$~ns per message). This is the \texttt{BQueueBatched}
mailbox of Section~\ref{sec:qtypes}. The gain is regime-dependent in exactly the
way the workload requires: it is negligible on a near-empty queue and maximal
when a burst has backed the queue up --- the same clustered-arrival regime
(Sections~\ref{sec:requirements},~\ref{sec:arrival}) that produces the latency
tail --- so batch draining is what holds the asynchronous stage fast under
precisely the load that matters. One result runs counter to intuition and fixes
where the batching must live: batching on the \emph{sender} side instead ---
pushing many messages under one lock --- \emph{lowered} throughput to $11.3$
million, because it serialized producer and consumer and destroyed their pipeline
overlap; fewer lock acquisitions did not compensate. Batching therefore belongs
on the drain side only.

We turn now to direct microbenchmarks of the synchronous path itself.

\subsection{Microbenchmarks of the C++ implementation}
\label{sec:micro}

Table~\ref{tab:elim} predicts that eliminating each per-message operation
produces a measurable reduction in latency. The framework's C++ microbenchmark
suite measures these steps on a \emph{sequential} round trip --- ping $\to$ pong
$\to$ reply with a single message in flight, so each sample is a latency rather
than a saturated throughput \cite{kasparperf}. The figures below were taken on an
AMD~EPYC~9374F (RHEL~9, \texttt{g++}~15), compiled \texttt{-O3 -march=native}
and pinned with \texttt{taskset} to two cores, reported as the median of three
runs. Absolute values depend on hardware, allocator, thread placement, and turbo
state; the reproducible result is the \emph{ratio} between rows, each of which
isolates one eliminated operation. The \texttt{steady\_clock} resolution on this
machine quantizes to $\sim\!10$~ns, so a per-sample \fs{} percentile is
resolution-bound; \fs{} costs are therefore also reported as amortized ns (total
wall time over $N$ operations, a single clock-read pair), which resolves costs
below one tick.

Table~\ref{tab:transport} isolates the cost of crossing threads; the two
paragraphs that follow it decompose the internal overhead of \fs{} and the
effect of the memory pool.

\paragraph{The cost is thread-crossing, not messaging.} Table~\ref{tab:transport}
holds the work constant --- one ping/pong round trip with a reply --- and varies
only the delivery mechanism, so each row adds back one operation. A cross-thread
asynchronous \snd{} between two actors on separate cores costs $\approx 3.4\,\mu$s
at the median. This cost is dominated by the two cross-core mailbox wakeups per
round trip, each a mutex acquisition and a condition-variable signal on which the
receiving thread must be scheduled; it is placement-sensitive, because the
cross-core path depends on which cores the two threads occupy. Keeping both
threads but letting each busy-poll its mailbox instead of sleeping removes the
wakeups and nothing else: the round trip falls to $370$~ns. Of the
$\sim\!3.4~\mu$s, therefore, some $3~\mu$s is putting threads to sleep and
waking them, and about $370$~ns --- two crossings of $\sim\!185$~ns --- is moving
the message and the queue slot between cores; the price of spinning is a core per
actor, busy whether or not a message arrives. Co-scheduling the two actors on one
thread (grouping) removes the crossings as well and reduces the round trip to
$90$~ns, a $37\times$ reduction. Replacing the remaining mailbox push/pop
with an inline \fs{} reduces the round trip to $\sim\!30$~ns (amortized; the
per-sample $p50$ is resolution-bound), because a \fs{} is mechanically a locked
function call: it takes the receiver's lock and runs the handler inline. This
$\sim\!30$~ns is the whole round trip --- the ping and pong invocations plus the
reply. What that costs against the dispatch a programmer would otherwise write by
hand is measured separately, in Table~\ref{tab:dispatch}.

\begin{table}[!htb]
\centering
\caption{\textbf{The cost is thread-crossing, not messaging.} Same ping/pong
round trip with a reply; only the delivery mechanism varies. Each row eliminates
one operation relative to the row above. ``Wakeup'' is a cross-core mailbox
wakeup of a sleeping thread (mutex $+$ condition variable); ``queue'' is a mailbox push/pop.
The $p50$ column is per-sample round-trip latency; the \fs{} figure is
amortized, because its per-sample $p50$ is resolution-bound (the
\texttt{steady\_clock} tick is $\sim\!10$~ns), and its speedup is computed on
amortized figures. The grouped and \fs{} rows show negligible spread. In the
spinning row both actors keep their own threads on separate cores, but each
thread busy-polls its mailbox (the lock-free mailbox with a consumer that never
sleeps), so a message still crosses cores but never waits for a thread to wake.
Three runs: $p50$ $370$--$380$~ns (median $370$), $p99$ $400$--$410$~ns.}
\label{tab:transport}
\small
\begin{tabular}{lccccr}
\toprule
Delivery mode & Threads & Wakeup & Queue & $p50$ RT (ns) & Speedup \\
\midrule
\snd{}, ungrouped        & 2 & yes & yes & $\sim\!3370$  & $1\times$ \\
\snd{}, ungrouped, spinning & 2 & no & yes & $370$ & $9.1\times$ \\
\snd{}, grouped          & 1 & no  & yes & $90$          & $37\times$ \\
\fs{} (inline)           & 1 & no  & no  & $\sim\!30$    & $\sim\!110\times$ \\
\bottomrule
\end{tabular}
\end{table}

\paragraph{Against hand-written dispatch.} Table~\ref{tab:transport} compares actor
delivery paths with each other. The comparison that decides whether to use the
framework at all is against the call a programmer would otherwise write.
Table~\ref{tab:dispatch} measures that on the same machine: every arm does the
same work on a stack-allocated message and differs only in how the work is
reached. It measures dispatch alone, without a reply, so its \fs{} figure is
below the round trip of Table~\ref{tab:transport}. Direct and virtual calls are
real, non-inlined calls. A virtual call costs very little when the branch
predictor knows its target and several nanoseconds when it does not, so the table
reports both: a single ``virtual call costs $X$'' figure would depend entirely on
how predictable the call site is.

\begin{table}[!htb]
\centering
\caption{\textbf{\fs{} against hand-written dispatch.} Same work in every arm;
amortized ns per call, minimum of 15 runs of $5\times10^6$ calls, AMD EPYC 9374F,
RHEL 9.2, \texttt{g++ -O3 -march=native}, pinned to two cores. Dispatch only, no
reply. Source: \texttt{actors/cpp/perf/bench\_dispatch.cpp}.}
\label{tab:dispatch}
\small
\begin{tabular}{lrr}
\toprule
Dispatch & ns & $\times$ direct \\
\midrule
direct call (not inlined)                         & $1.63$ & $1.0$ \\
virtual call, one target (predicted)             & $1.86$ & $1.1$ \\
virtual call, 2--8 targets (mispredicted)        & $7.53$--$9.32$ & $4.6$--$5.7$ \\
\fs{}                                            & $9.10$ & $5.6$ \\
\bottomrule
\end{tabular}
\end{table}

\fs{} costs about one virtual call whose target the predictor cannot guess, and
about $5\times$ a predicted virtual call or a direct call: some $7.5$~ns more per
hop, for the receiver's lock, message-field setup, handler lookup and reply
plumbing. Whether that is small depends on what it is compared with. Against the
$\sim\!1~\mu$s one-thread socket-to-book median of Section~\ref{sec:casestudy}
it is about $1\%$; on a leaner path --- a hardware-timestamped, wire-to-book pipeline measured
in hundreds of nanoseconds --- the same $7.5$~ns per hop would matter.

\paragraph{Grouping, and why \fs{} improves on it.} The grouped row is the
strongest conventional alternative. Grouping (Section~\ref{sec:groups})
co-schedules the two actors on one thread behind one shared mailbox, removing
both cross-core wakeups and reducing the round trip from $\sim\!3.4\,\mu$s
to $90$~ns. Grouping still pays for the queue: the message is enqueued, the run
loop dequeues it, and the handler runs on a later turn. \fs{} removes that layer.
On the same thread it does not enqueue; it takes the receiver's lock and runs the
handler inline, returning the reply as a value, in $\sim\!30$~ns --- about
$3\times$ faster than grouping. Grouping removes the thread hop; \fs{} removes the
thread hop \emph{and} the queue. This synchronous, inline, receiver-transparent
delivery is, to our knowledge, absent from the mainstream actor systems surveyed
in Section~\ref{sec:survey}. The intended use follows: group the actors of a
servicing stage onto one thread and drive them with \fs{}, so that an $N$-actor
pipeline collapses to a single inline call chain (the middle section of
Figure~\ref{fig:hftarch}).

\paragraph{Allocation, not dispatch, is the only variable cost.} Beyond the
inline dispatch, the only per-message cost a caller pays is for memory it chooses
to allocate. A request message kept on the stack requires none, because \fs{}
blocks until the handler returns (Section~\ref{sec:mech}). For a message that must
outlive the call, allocation from the \texttt{MemoryPool} of
Section~\ref{sec:pool} replaces a general-purpose
\texttt{new}/\texttt{delete}. On a grouped-\snd{} round trip with two allocations,
the pool reduces the amortized cost from $\sim\!128$~ns (pool off) to $\sim\!92$~ns
(pool on). The larger and platform-independent benefit is the tail: in a separate
run the maximum round trip fell from $5.5$~ms to $33\,\mu$s once pooled, because
the pool never takes the general allocator's occasional slow path: a
two-order-of-magnitude reduction of the $p99.9{+}$ tail that an unpooled
servicing stage would otherwise inject into the path.

\paragraph{Scope.} These are sequential latencies with one message in flight, not
saturated throughput, and they do not vary contention or chain depth. The
established results are: \fs{} costs tens of nanoseconds; its overhead over a
direct call is $\sim\!7.5$~ns plus whatever memory the caller allocates; it
improves on same-thread grouping by about $3\times$; and pooling removes the
latency tail. A saturated, multi-axis evaluation
remains future work.

\subsection{Mailbox queue types and their selection}
\label{sec:qtypes}

The four queue implementations of Section~\ref{sec:queues-mech} ---
\texttt{BQueue}, \texttt{BQueueBatched}, \texttt{ShardedBQueue}, and
\texttt{LockFreeMPSC} --- separate on three regimes, which
Table~\ref{tab:qtypes} measures on the EPYC target. A fourth regime, a single
cross-thread message, distinguishes none of them, because thread placement moves
the figure by $2.2\times$ --- more than the gap between queues.

\begin{table}[!htb]
\centering
\caption{\textbf{Four mailbox queue types, one column per distinguishing regime}
(EPYC target, pinned, median of three runs). \emph{Grouped}: single-thread
round-trip median. \emph{Burst}: single bursty producer, amortized cost per
message. \emph{Fan-in}: 32 concurrent producers, push $p99$. Bold marks the best
in each regime; no type is best in all three.}
\label{tab:qtypes}
\small
\begin{tabular}{lrrr}
\toprule
Queue type & Grouped $p50$ (ns) & Burst (ns/msg) & Fan-in $p99$ ($\mu$s) \\
\midrule
\texttt{BQueue}        & $90$          & $605.8$          & $69.9$ \\
\texttt{BQueueBatched} & $90$          & $\mathbf{584.9}$ & $72.4$ \\
\texttt{ShardedBQueue} & $190$         & $983.1$          & $\mathbf{6.6}$ \\
\texttt{LockFreeMPSC}  & $\mathbf{70}$ & $672.0$          & $16.5$ \\
\bottomrule
\end{tabular}
\end{table}

No single queue wins everywhere. On a shared (grouped) thread, \texttt{LockFreeMPSC}
has the lowest median ($70$~ns, $1.3\times$ below \texttt{BQueue}), because its
push and pop are each an uncontended atomic operation with no lock or wakeup.
Under a bursty single producer --- the market-data regime of the case study ---
\texttt{BQueueBatched} is fastest ($584.9$~ns/msg): draining an entire packet
burst under one lock acquisition amortizes the lock over the burst, the
drain-side batching whose throughput effect is decomposed in
Section~\ref{sec:micro}, and
\texttt{BQueue} trails by under $4\%$. Under 32-producer fan-in, \texttt{ShardedBQueue}
wins decisively --- $p99$ of $6.6~\mu$s against \texttt{BQueue}'s $69.9~\mu$s
($10.6\times$), and $3.6\times$ the throughput --- because its independent lanes
remove the single-mutex serialization that dominates the other three; it pays for
those lanes in the other regimes, where it is the slowest of the four.

A selection heuristic for the tested EPYC configuration follows; it is a measured
guideline for this hardware and these regimes, not a general rule. \texttt{BQueue}
is a reasonable default: it is competitive in the grouped and bursty regimes
(within $\sim\!30\%$ of \texttt{LockFreeMPSC} grouped, within $4\%$ of
\texttt{BQueueBatched} bursty) and has the most stable tail of the four (grouped
\texttt{max} $2.4$--$9~\mu$s, versus $14$--$16~\mu$s for \texttt{LockFreeMPSC}).
\texttt{BQueueBatched} is preferable for bursty single-producer mailboxes such as
a market-data ingest, and \texttt{ShardedBQueue} for the few actors written by
many producers concurrently --- a shared order-out actor or a fan-in aggregator
--- the one regime in which it wins decisively. \texttt{LockFreeMPSC} has the best
grouped median but the least stable tail measured, so we do not default to it; and
\texttt{ShardedBQueue}'s lanes are wasted, and counterproductive, outside fan-in.

\FloatBarrier

% ---------------------------------------------------------------
\section{Case study: tick-to-book latency in a live market-data path}
\label{sec:casestudy}
% ---------------------------------------------------------------

The microbenchmarks of Section~\ref{sec:micro} measure the framework's
per-message cost in isolation. This
section grounds that number in a live measurement of the \textsc{kaspar-hft}
market-data path on production exchange data, and asks the paper's title question
directly: is the actor model's overhead small enough for a high-frequency
market-data path? We measured the deployed system in two configurations. In the
\emph{one-thread} configuration, the socket-reader thread decodes each datagram
and updates the book itself. In the \emph{two-stage} configuration, the reader
hands each datagram to a decode-and-book actor on a second thread through an
asynchronous mailbox. The one-thread path is the faster of the two and provides
the main result; the two-stage path is reported for comparison. Because its
mailbox makes queue occupancy observable, the two-stage path also admits a finer
decomposition of latency into a floor, an in-packet serialization cost, and a
queueing component, which Sections~\ref{sec:floorslope}--\ref{sec:queue}
develop.

\subsection{Setup}

The path under measurement is the market-data ingest of Section~\ref{sec:hftarch}:
read a UDP datagram off CME MDP~3.0 multicast, decode the Simple Binary Encoding
(SBE) messages \cite{fixsbe} inside it, apply them to an order book, and publish.
Two timestamps are taken per message --- $t_0$ when the datagram comes up from the
socket, $t_1$ when the book update is published --- and their difference is
recorded for every message. Because $t_0$ is a \emph{software} timestamp taken at
the socket read, time spent in the NIC, the kernel, and the socket buffer is
excluded: the measurement is socket-to-book, not wire-to-book.

On the two-stage path the reader passes each datagram to the decode-and-book
actor by an asynchronous \snd{}. The send is necessarily asynchronous, since a
blocking reader would stop draining its socket and drop packets. Decode and book
construction then run together on the second thread. The hand-off between the
two stages is the decoder's mailbox, a bounded ring buffer of datagrams (a
\texttt{BQueue}), and we write $\text{qlen}$ for its occupancy: the number of
datagrams already waiting in the ring when a given datagram is handed to the
decoder, so that $\text{qlen}=0$ means the ring was empty and the datagram was
picked up at once. References below to the ring or the queue are to this
mailbox. The hand-off has a single bursty producer, the regime in which
\texttt{BQueueBatched} is marginally faster than the \texttt{BQueue} used here
(Section~\ref{sec:qtypes}); the difference is a few percent of the mailbox cost
and negligible against the $\sim\!7~\mu$s floor of this path. The one-thread path
has no hand-off, no ring, and no $\text{qlen}$.

Both paths were measured on live CME futures for three instruments --- the
E-mini S\&P~500 (ES), the E-mini Nasdaq-100 (NQ), and the 10-year Treasury Note
(ZN) --- each with a book-update and a trade stream. Latency on the one-thread
path was measured in two five-minute runs, discarding the first minute of each;
the receiver used kernel-bypass networking (Onload) on a single feed, with
feed~B switched off, and threads were not pinned to cores. The two-stage data
comes from a single $53$-minute session on a busy afternoon, $8.29$~million
messages in all.

\subsection{The one-thread path}
\label{sec:onethread}

Table~\ref{tab:onethread} gives the distribution of socket-to-book latency on
the one-thread path. The book-update medians are $0.8$--$1.1~\mu$s. This is the
cost of decode and book update themselves, the work every message pays when no
cross-thread hand-off lies in its path. Against it, a \fs{} hop ($\sim\!9$~ns,
Table~\ref{tab:dispatch}) is about $1\%$.

\begin{table}[!htb]
\centering
\caption{Socket-to-book latency ($\mu$s) on the one-thread path, by stream.}
\label{tab:onethread}
\small
\begin{tabular}{lrrrrrrr}
\toprule
Stream & Messages & $p1$ & $p10$ & $p50$ & $p90$ & $p99$ & $p99.9$ \\
\midrule
ES book  & 401{,}462 & 0.5 & 0.7 & 1.1 & 2.7 & 9.9 & 24.1 \\
NQ book  & 581{,}010 & 0.4 & 0.5 & 0.8 & 1.9 & 4.7 & 8.8 \\
ZN book  & 141{,}972 & 0.4 & 0.6 & 1.0 & 2.5 & 20.6 & 69.3 \\
ES trade & 45{,}733 & 0.5 & 0.7 & 2.5 & 8.8 & 36.2 & 70.1 \\
NQ trade & 21{,}093 & 0.4 & 0.6 & 0.9 & 3.3 & 17.8 & 31.5 \\
ZN trade & 13{,}282 & 0.4 & 0.7 & 6.6 & 41.5 & 90.5 & 124.0 \\
\bottomrule
\end{tabular}
\end{table}

\paragraph{Comparison with the two-stage path.} On the two-stage path the
book-update medians are $7.1$--$7.3~\mu$s (Table~\ref{tab:mddist}), about
$6~\mu$s higher. The main structural difference between the two paths is the
cross-thread hand-off through the asynchronous mailbox, and the direction is the
one the paper's own measurements predict: waking a sleeping thread costs
microseconds (Table~\ref{tab:transport}). We do not, however, attribute the
whole $\sim\!6~\mu$s to the hand-off, because the runs also differ in market
load, in length, and possibly in network stack and configuration. The floor and
tail reported for the two-stage path below are therefore properties of that
path, not the cost of decode and book update.

\subsection{The per-message model: floor plus serialization}
\label{sec:floorslope}

Latency on both paths follows the same per-message model; what differs is the
floor. The decode-and-book work itself sets a floor of about $1~\mu$s, as the
one-thread path shows (Section~\ref{sec:onethread}). On the two-stage path the
floor is higher, $\sim\!7~\mu$s, because each datagram also crosses a thread
through the mailbox, and because of the other differences between the runs noted
above. The coefficients and the serialization evidence below come from the
two-stage run, for which each message's in-packet position and queue occupancy
were recorded.

One UDP datagram carries many SBE messages, decoded in order. Write $\text{idx}$
for a message's position within its packet, counted from zero: $\text{idx}=0$ is
the message decoded first, $\text{idx}=1$ the next, and so on. A message at
position $\text{idx}$ therefore waits for the $\text{idx}$ decodes ahead of it
before its own runs --- a wait that sits inside its measured latency, and that is
zero exactly when $\text{idx}=0$. Restricting to messages that arrive to an empty
ring (nothing queued ahead of their packet), so the only remaining wait is
in-packet, the median latency is linear in $\text{idx}$,
\[
  \text{median latency} \;\approx\; \text{floor} \;+\; \text{slope}\times\text{idx},
\]
with the two coefficients of Table~\ref{tab:floorslope}, fitted on medians. (The
mean is unsuitable here: a single multi-millisecond stall in a cell moves the mean by
tens of microseconds, so every quantity in this section is a median or a
percentile.)

The \emph{floor} --- the intercept, a message first in its packet --- agrees to
within $0.42~\mu$s across six independent streams, $6.8$--$7.2~\mu$s. It
comprises the $\sim\!1~\mu$s of decode-and-book work together with the
cross-thread hand-off and whatever else differs between the runs; it is not the
cost of decode and book alone. It is insensitive to queue length, and it is
stable under
load, moving by at most $+0.4~\mu$s (and \emph{falling} on two of three
instruments) during a Federal Reserve release that raised the packet rate
$4.6$--$9.8\times$ in a second. The synchronous per-hop cost is small against
either floor: a \fs{} hop adds $\sim\!7.5$~ns over a direct call ($\sim\!9$~ns in
all, about one mispredicted virtual call, Table~\ref{tab:dispatch}; the full
ping/pong round trip is $\sim\!30$~ns), about $1\%$ of the one-thread median. The
asynchronous cross-thread hand-off is not small: it contributes to the floor of
the two-stage path, and its queue to that path's tail (Section~\ref{sec:queue}).

The \emph{slope} --- the per-message service cost within the decode-and-book
stage, $0.31$--$0.97~\mu$s; whether it is dominated by SBE decode, by book
mutation, or by cache effects we do not resolve here --- is the dominant measured
contributor to the within-packet component of the tail. In-packet position scales
linearly into latency: the last
message of a $45$-message packet incurs $44\times0.57 \approx 25~\mu$s on top of
the floor, deterministically. On the trade streams a message is essentially never alone
($99.3$--$99.6\%$ have another message ahead of them in the same packet), which is
why the trade medians already sit above the book medians. Shaving $100$~ns off the
floor moves every message by $100$~ns; shaving it off the slope moves a message by
$\text{idx}\times100$~ns, and $\text{idx}$ is largest in the tail. On the
two-stage path the design lever for the tail is therefore the slope rather than
the mailbox queue, which is rarely occupied: between $87\%$ and $99.7\%$ of
messages arrive to find the ring empty. The floor itself is lowered by removing
the hand-off, as the $0.8$--$1.1~\mu$s book-update medians of the one-thread path
show (Table~\ref{tab:onethread}).

\begin{table}[t]
\centering
\caption{The per-message model
$\text{median}\approx\text{floor}+\text{slope}\times\text{idx}$ on the two-stage
path, fitted on medians at queue length zero. The floor agrees within
$0.42~\mu$s across six streams; the slope, the per-message decode-and-book cost,
varies $\sim\!3\times$ and drives the within-packet tail.}
\label{tab:floorslope}
\small
\begin{tabular}{lrr}
\toprule
Stream & Floor ($\mu$s) & Slope (ns/msg) \\
\midrule
ES book  & 6.98 & 566 \\
NQ book  & 7.23 & 312 \\
ZN book  & 6.83 & 966 \\
ES trade & 6.81 & 714 \\
NQ trade & 7.14 & 526 \\
ZN trade & 6.81 & 965 \\
\bottomrule
\end{tabular}
\end{table}

\begin{figure}[t]
\centering
\begin{tikzpicture}
\begin{axis}[
  width=0.68\linewidth, height=6.4cm,
  xlabel={In-packet position idx}, ylabel={Median latency ($\mu$s)},
  domain=0:40, samples=2,
  xmin=0, xmax=40, ymin=6,
  legend style={at={(1.02,1)},anchor=north west,font=\small},
  legend cell align=left,
  grid=both, grid style={gray!20},
]
\addplot[blue,thick]           {6.98+0.566*x}; \addlegendentry{ES book (566 ns/msg)}
\addplot[red,thick]            {7.23+0.312*x}; \addlegendentry{NQ book (312)}
\addplot[green!55!black,thick] {6.83+0.966*x}; \addlegendentry{ZN book (966)}
\addplot[blue,dashed]          {6.81+0.714*x}; \addlegendentry{ES trade (714)}
\addplot[red,dashed]           {7.14+0.526*x}; \addlegendentry{NQ trade (526)}
\addplot[green!55!black,dashed] {6.81+0.965*x}; \addlegendentry{ZN trade (965)}
\end{axis}
\end{tikzpicture}
\caption{The per-message model $\text{median}=\text{floor}+\text{slope}\times
\text{idx}$ (Table~\ref{tab:floorslope}), one line per stream. All six share a $\sim\!7~\mu$s floor at $\text{idx}=0$; the slope ---
the per-message serialization cost --- sets the divergence, from $312$~ns/msg (NQ~book) to
$966$~ns/msg (ZN~book).}
\label{fig:floorslope}
\end{figure}
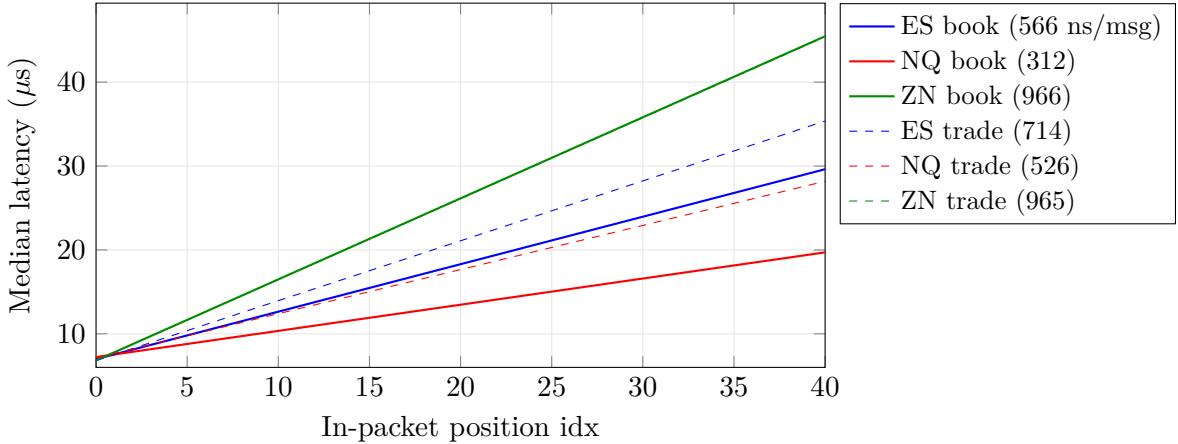

\subsection{Three estimates of the floor}
\label{sec:floor-robust}

The floor of Table~\ref{tab:floorslope} is a fitted intercept. Two further
estimates, constructed differently, agree with it. The first isolates the hot path
directly: the subset of messages with $\text{qlen}=0$ and $\text{idx}=0$, for which
neither the queue nor in-packet serialization contributes, so its median is the
floor with no model interposed (Table~\ref{tab:floordirect}). Across six streams
the spread is $1.41~\mu$s; across the three book streams it is $0.35~\mu$s.

\begin{table}[!htb]
\centering
\caption{Direct estimate of the floor: median latency ($\mu$s) and sample size
of the messages at $\text{qlen}=0$ and $\text{idx}=0$, for which neither
queueing nor in-packet serialization contributes.}
\label{tab:floordirect}
\small
\begin{tabular}{lrr}
\toprule
Stream & Median ($\mu$s) & $n$ \\
\midrule
ES book  & 7.01 & 1{,}781{,}767 \\
NQ book  & 7.24 & 1{,}841{,}447 \\
ZN book  & 6.89 & \phantom{0,}671{,}271 \\
ES trade & 7.62 & \phantom{0,00}1{,}008 \\
NQ trade & 8.30 & \phantom{0,000,}298 \\
ZN trade & 7.96 & \phantom{0,000,}474 \\
\bottomrule
\end{tabular}
\end{table}

The second estimate is the fitted $\text{idx}=0$ intercept of the median ladder
(Table~\ref{tab:floorslope}). The third is measured during the Federal Open Market
Committee statement, when the packet rate rose $6.5\times$, $4.6\times$, and
$9.8\times$ on the three instruments within a second; taking the floor from
single-message packets ($\text{span}\approx1$) in that window gives the third row
of Table~\ref{tab:floor3}. The three agree to within a few tenths of a
microsecond.

\begin{table}[!htb]
\centering
\caption{Three independent estimates of the floor ($\mu$s) for the book
streams. The three methods differ in their sample, their estimator, and the load regime
they were taken under, and agree to within $0.4~\mu$s.}
\label{tab:floor3}
\small
\begin{tabular}{lrrr}
\toprule
Method & ES & NQ & ZN \\
\midrule
Direct median ($\text{qlen}=0$, $\text{idx}=0$)        & 7.01 & 7.24 & 6.89 \\
Fitted $\text{idx}=0$ intercept (median ladder)        & 6.98 & 7.23 & 6.83 \\
Floor at $\text{span}\approx 1$ during the FOMC release & 7.1\phantom{0} & 7.5\phantom{0} & 7.0\phantom{0} \\
\bottomrule
\end{tabular}
\end{table}

The third estimate is the most informative. During the release the floor moved
by at most $+0.4~\mu$s, and fell on two of the three instruments. The hot path is
not rate-sensitive: three methods, three instruments, all within
$6.8$--$7.5~\mu$s. The floor is the most stable quantity in the two-stage
measurements --- it moved negligibly when the
sample tripled, when the estimator changed from mean to median, and when the
packet rate rose tenfold.

\subsection{The distribution and its tail}

For a trading system the relevant quantity is the tail, not the median.
Table~\ref{tab:mddist} gives the unconditional distribution on the two-stage
path. The medians cluster
--- three separate multicast channels, three instruments, all within $0.2~\mu$s
--- but the tails diverge: ES book's $p99$ is $18.5~\mu$s, ZN book's is
$57.0~\mu$s, and ZN trade's is $219~\mu$s. Expressed as amplification from median
to $p99$, the streams range from $1.9\times$ (NQ book) to $14.2\times$ (ZN
trade). A single ``$8~\mu$s'' figure would be accurate for the median but
uninformative about the tail. The one-thread path shows the same divergence one
level down (Table~\ref{tab:onethread}): book-update medians of
$0.8$--$1.1~\mu$s, but a ZN~trade $p99$ of $90.5~\mu$s.

\begin{table}[t]
\centering
\caption{Unconditional socket-to-book latency distribution ($\mu$s) on the
two-stage path, over every admitted message of each stream. The median is
uniform across streams; the tail is not.}
\label{tab:mddist}
\small
\begin{tabular}{lrrrrrr}
\toprule
Stream & Messages & $p50$ & $p90$ & $p99$ & $p99.9$ & max \\
\midrule
ES book  & 2{,}861{,}519 & 7.1 & 10.6 & 18.5 & 41.2 & 1148.4 \\
NQ book  & 3{,}792{,}033 & 7.3 & \phantom{0}9.9 & 13.6 & 24.7 & 5567.0 \\
ZN book  & 1{,}119{,}746 & 7.1 & 12.1 & 57.0 & 180.9 & 2458.2 \\
ES trade & \phantom{0,}296{,}736 & 9.0 & 15.3 & 38.0 & 127.7 & \phantom{0}421.5 \\
NQ trade & \phantom{0,}110{,}003 & 8.3 & 12.5 & 31.2 & 83.7 & \phantom{0}694.2 \\
ZN trade & \phantom{0,}114{,}154 & 15.4 & 69.3 & 219.4 & 409.5 & \phantom{0}504.8 \\
\bottomrule
\end{tabular}
\end{table}

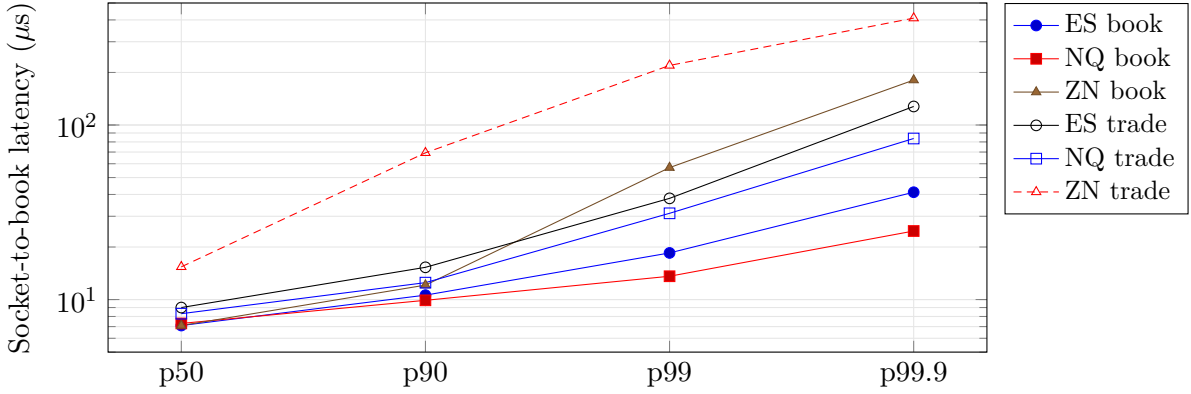
\begin{figure}[t]
\centering
\begin{tikzpicture}
\begin{axis}[
  width=0.8\linewidth, height=6.2cm,
  ymode=log, log basis y=10,
  symbolic x coords={p50,p90,p99,p99.9},
  xtick=data, ymin=5, ymax=500,
  ylabel={Socket-to-book latency ($\mu$s)},
  legend style={at={(1.02,1)},anchor=north west,font=\small},
  legend cell align=left, enlarge x limits=0.1,
  grid=both, grid style={gray!20},
]
\addplot+[mark=*]         coordinates {(p50,7.1)(p90,10.6)(p99,18.5)(p99.9,41.2)};  \addlegendentry{ES book}
\addplot+[mark=square*]   coordinates {(p50,7.3)(p90,9.9)(p99,13.6)(p99.9,24.7)};   \addlegendentry{NQ book}
\addplot+[mark=triangle*] coordinates {(p50,7.1)(p90,12.1)(p99,57.0)(p99.9,180.9)}; \addlegendentry{ZN book}
\addplot+[mark=o]         coordinates {(p50,9.0)(p90,15.3)(p99,38.0)(p99.9,127.7)}; \addlegendentry{ES trade}
\addplot+[mark=square]    coordinates {(p50,8.3)(p90,12.5)(p99,31.2)(p99.9,83.7)};  \addlegendentry{NQ trade}
\addplot+[mark=triangle]  coordinates {(p50,15.4)(p90,69.3)(p99,219.4)(p99.9,409.5)}; \addlegendentry{ZN trade}
\end{axis}
\end{tikzpicture}
\caption{Socket-to-book latency percentiles per stream
(Table~\ref{tab:mddist}), log scale. Medians cluster near
$7$--$15~\mu$s; the tails diverge, ZN~trade
reaching $219~\mu$s at $p99$ and $410~\mu$s at $p99.9$.}
\label{fig:mddist}
\end{figure}

Two mechanisms produce this tail, and the next two subsections take them in turn:
queue occupancy, when a burst backs up the decoder's ring
(Section~\ref{sec:queue}), and in-packet position, when a datagram carries many
messages (Section~\ref{sec:suitability}). The first exists only because the
two-stage path hands datagrams across threads through a mailbox. Their
full deconfounding, and how to mitigate the tail, is in a companion
paper~\cite{mdarrival}.

\subsection{Latency versus queue occupancy}
\label{sec:queue}

\begin{figure}[!htb]
\centering
\begin{tikzpicture}
\begin{axis}[
  width=0.72\linewidth, height=6cm,
  xlabel={queue occupancy $\text{qlen}$ (packets waiting)},
  ylabel={median latency ($\mu$s, $\text{idx}=0$)},
  xmin=0, xmax=6, ymin=5,
  xtick={0,1,2,3,4,5,6},
  legend style={at={(0.02,0.98)},anchor=north west,font=\small},
  legend cell align=left,
  grid=both, grid style={gray!20},
  every axis plot/.append style={thick},
]
\addplot[blue,mark=*]              coordinates {(0,6.9)(1,6.3)(2,7.1)(3,8.5)(4,9.8)(5,11.5)(6,16.6)}; \addlegendentry{ES book}
\addplot[red,mark=square*]         coordinates {(0,7.2)(1,5.9)(2,6.9)(3,8.3)(4,11.2)(5,16.4)};        \addlegendentry{NQ book}
\addplot[green!55!black,mark=triangle*] coordinates {(0,6.9)(1,6.0)(2,7.6)(3,9.4)(4,11.6)(5,21.7)};   \addlegendentry{ZN book}
\end{axis}
\end{tikzpicture}
\caption{\textbf{Latency rises with queue occupancy.} Median socket-to-book latency against queue occupancy $\text{qlen}$ for the three book streams, with
in-packet position held at $\text{idx}=0$ to isolate the queue from packet size.
At $\text{qlen}=0$ --- the state $87\%$--$99.7\%$ of messages find --- the median
sits at the $\sim\!7~\mu$s floor; each further queued packet adds more than the
last (convex), so the tail, where it is queue-driven, tracks $\text{qlen}$. The
effect is confined to a rare event ($\text{qlen}\ge3$ on $0.08\%$ of messages).
Deep-queue points ($\text{qlen}\ge5$) rest on few messages and are noisier.}
\label{fig:convex}
\end{figure}
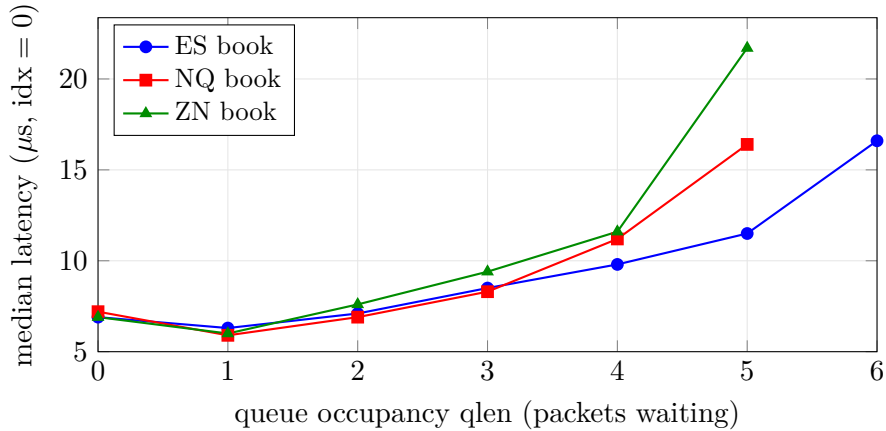

Recall that $\text{qlen}$ is the number of datagrams already waiting in the
decoder's ring when a datagram is handed to it, so that $\text{qlen}=0$ means an
empty ring. Figure~\ref{fig:convex} plots the median
socket-to-book latency at each queue length for the three book streams, with
in-packet position held at $\text{idx}=0$ so the curve isolates the queue from
packet size. Two features stand out.

\emph{At $\text{qlen}=0$ the path is at its floor.} The median sits at the
$\sim\!7~\mu$s baseline of Section~\ref{sec:floorslope}, and this is the state
$87\%$ to $99.7\%$ of messages find: an empty ring adds no queueing wait. For the
overwhelming majority of messages the path runs at that baseline.

\emph{As the ring fills, latency rises, and convexly.} Each additional queued
packet costs more than the last: ES~book climbs from $6.9~\mu$s at $\text{qlen}=0$
to $16.6~\mu$s at $\text{qlen}=6$, and ZN~book from $6.9$ to $21.7~\mu$s at
$\text{qlen}=5$. This is ordinary queueing --- when a burst arrives faster than the
decoder drains it, packets wait behind packets, and the wait grows with occupancy
--- and it is why the latency tail of the two-stage path is, in part, a
queue-occupancy effect of the cross-thread hand-off. It is a \emph{rare} effect: $\text{qlen}\ge3$ occurs on only $0.08\%$ of ES~book messages
($2{,}311$ of $2{,}861{,}519$), so the queue is a large effect at the far
percentiles on a small fraction of messages, not a small effect everywhere. Its
full deconfounding from packet size is carried out in the companion paper
\cite{mdarrival}.

\subsection{Suitability: the per-hop cost and the hand-off}
\label{sec:suitability}
\label{sec:arrival}

The framework enters the measured latency in two different ways, and they
differ by orders of magnitude. The synchronous per-hop cost is small: one \fs{}
dispatch costs about one mispredicted virtual call ($\sim\!9$~ns,
Table~\ref{tab:dispatch}), about $1\%$ of the $\sim\!1~\mu$s book-update median
of the one-thread path (Section~\ref{sec:onethread}) and about $0.1\%$ of the
$\sim\!7~\mu$s floor of the two-stage path. It does not grow with in-packet
position. The asynchronous cross-thread hand-off is not small. It is the main
structural difference between the $\sim\!7~\mu$s medians of the two-stage path
and the $0.8$--$1.1~\mu$s medians of the one-thread path
(Section~\ref{sec:onethread}), so it contributes to the floor, and through its
ring it contributes to the tail (Section~\ref{sec:queue}).

The tail of the two-stage path has two sources. First, \emph{queue occupancy}: when a burst
backs the ring up, latency rises with $\text{qlen}$ (Section~\ref{sec:queue})
--- a genuine tail contributor, but a rare one, since $87\%$ to $99.7\%$ of
messages find an empty ring; it exists because of the cross-thread hand-off.
Second, and for most messages the larger effect, \emph{in-packet position}: a
datagram carries several SBE messages decoded in sequence, so a message late in a
large packet waits behind the earlier ones. In-packet position is a property of
the feed --- bursts of market events that the exchange coalesces into large
packets --- and applies on either path. The full characterization of
the arrival process, and what it implies for mitigating the tail, is the subject
of a companion paper \cite{mdarrival}.

The verdict is affirmative and specific, and it is the paper's case for \fs{}
and groups: the per-hop cost of synchronous delivery is about $1\%$ of the
decode-and-book work, while an asynchronous cross-thread hand-off in the path
raises the floor and adds a queueing tail. Keeping the path on one thread brings
the book-update median to about $1~\mu$s. What an implementer must engineer is
where the path crosses threads, the per-message service time --- the
decode-and-book slope --- and the feed's batch structure.

\subsection{Limits of these measurements}

Because $t_0$ is a software timestamp at the socket read, both sets of
measurements are socket-to-book rather than wire-to-book; hardware receive
timestamps would close that gap and were not taken. The two-stage data comes
from one machine and one $53$-minute session, and the one-thread data from two
short runs in a separate session, on a single feed with threads not pinned.
Because the two paths were not measured side by side, the $\sim\!6~\mu$s between
their medians is not attributed to the hand-off alone. On the two-stage path the
per-message service cost varies $\sim\!3\times$
across instruments (ZN is dearest), and whether that is cold-cache residency or a
genuinely heavier message mix cannot be separated without hardware counters. And a
separate population of multi-millisecond stalls --- the largest latencies in the
two-stage data (the \texttt{max} column of Table~\ref{tab:mddist}) --- is real and
remains unexplained here.

% ---------------------------------------------------------------
\section{Production and simulation: a duality of the group}
\label{sec:duality}
% ---------------------------------------------------------------

\paragraph{The group defines a single message order.} A \texttt{Group}
co-schedules its member actors on one thread behind one shared mailbox
(Section~\ref{sec:micro}): the members do not own separate queues; the group's
single queue holds every message addressed to any member, and the run loop
dispatches them one at a time. Beyond the latency effect measured earlier --- the
collapse of the round trip from $\sim\!3.4\,\mu$s to $90$~ns --- this has a
structural consequence. A single queue imposes a single total order on all
messages entering the group, so the group's entire execution --- every actor
state transition and every outgoing message --- is a deterministic function of
that one input sequence.

\paragraph{The same actors run in production and in simulation.} That property
makes one group two systems. In production the queue is fed by live sources ---
socket readers, timers, market-data actors --- and the run loop executes in real
time. In simulation the queue is fed from a recorded or synthetic sequence, and
the run loop executes as fast as the host permits. The member actors are
identical in both modes: the framework gives a handler nothing that
distinguishes a message from a live socket from one read from a backtest file,
which is the receiver transparency of Section~\ref{sec:transparency} extended
from the delivery mechanism to the data source. As there, this is a property of
the interface, not isolation the language enforces. No strategy actor is edited, recompiled, or
conditionally branched to move between live trading and backtest. Given the same
ordered input sequence and deterministic injected services, replay reproduces the
same state transitions and outputs --- the single shared queue fixes the message
order, and the actors add no other source of nondeterminism. That is precisely
the one discipline the duality imposes: an actor must acquire time, randomness,
and external input through messages or injected interfaces rather than through
global calls, so that the simulator can supply them deterministically.

\paragraph{Relation to prior work.} Two ingredients of this design have prior
art; their combination is uncommon. The single-shared-queue group is a
first-class facility in Actor4j \cite{actor4j2018}, where a group's actors run on
one thread with their mailboxes outsourced to the thread's queue --- presented
there as a message-passing throughput optimization, not as a simulation
mechanism. The same-code production/simulation duality is the principle of
deterministic simulation testing: FoundationDB's Flow runtime \cite{foundationdb}
and systems such as TigerBeetle run unmodified production code inside a
single-threaded deterministic simulator, but there the unit of determinism is an
asynchronous continuation or an explicit state machine rather than a group of
mailbox actors sharing a queue. Agent-based market simulators such as ABIDES
\cite{abides} drive actor-like agents behind one time-ordered event queue, but
those agents exist only in simulation and are not a deployed production system.
The design here couples the two: the shared-queue group is what makes
mode-agnostic actors run identically in real-time production and in deterministic
backtest. We are not aware of this specific combination being described
elsewhere.

\paragraph{Example: developing an execution algorithm.} This duality is how the
order-level shadowing execution algorithm of the companion paper \cite{shadowpov}
was developed. The strategy is an actor that consumes market-data and order-event
messages and emits child-order messages; it was built and validated entirely in
simulation, driven from recorded CME sessions, and then deployed to production
without modification --- the same actor and the same handlers, reading from a
live feed instead of a file. The $246$-session slippage comparison reported there
was produced by the simulation mode of the same group that runs the strategy
live. The child-order logic is written once and is agnostic to the mode in which
it executes.

% ---------------------------------------------------------------
\section{Discussion and limitations}
\label{sec:disc}
% ---------------------------------------------------------------

\paragraph{The scope is co-located actors.} \fs{} applies only when sender and
receiver share an address space; a synchronous cross-machine call would
reintroduce the network round trip it exists to avoid, and the actor model's
location transparency means the sender can fall back to asynchronous delivery
when the receiver is remote. Receiver transparency is what makes that fallback
free: the same handler serves the remote asynchronous case and the local
synchronous case.

\paragraph{The cost of transparency.} Heap-allocating the reply in both modes is
a deliberate cost paid to keep the handler unable to distinguish them. The
optimized synchronous-only reply removes that allocation but forfeits
transparency and must therefore be confined to code paths statically known to be
synchronous; mixing it with asynchronous delivery is an error the mechanism
reports rather than tolerates.

\paragraph{Blocking runs counter to conventional guidance.} Established actor
frameworks discourage blocking inside actors precisely because it can starve
thread pools and cause deadlock \cite{akka,armstrong2003,swift2021}. \fs{}
blocks the sender by construction, and its safety rests on the cyclic-invocation
detector and on the availability of the non-blocking and best-effort variants for
paths that cannot tolerate an unbounded wait. Where a caller must never block,
\texttt{fast\_send\_x} and \texttt{fast\_send\_void} provide bounded-time and
never-blocking alternatives behind the same handler interface. Even so, the
call-chain detector addresses only cycles on one thread
(Section~\ref{sec:cycle}), and only \emph{cyclic} waits; avoiding
thread-pool starvation for acyclic blocking patterns --- a chain that blocks on
an actor whose handler is itself waiting on a scarce resource --- remains a
burden on the developer that asynchronous delivery does not impose.

\paragraph{Contention on a hot actor.} A synchronous send holds the receiver's
exclusive-access lock for the duration of handler execution, so concurrent
synchronous sends to the same heavily contended actor serialize on that lock, and
the blocking variants stall their caller threads while they wait. Under heavy
contention this can perform worse than asynchronous delivery, in which senders
enqueue and continue and the messages buffer without stalling the callers. The
non-blocking (\texttt{fast\_send\_x}) and best-effort (\texttt{fast\_send\_void})
variants bound or remove the stall --- the latter by falling back to the mailbox
--- so the ratio of synchronous completions to asynchronous fallbacks doubles as
a signal that a target has become hot enough that the asynchronous path is
preferable.

\paragraph{Permissive detection shifts a burden to the programmer.} The
(actor, message-class) refinement admits callbacks by allowing an actor's state
to be re-entered by a nested handler, which reintroduces the reentrancy hazards
that strict per-actor sequential processing removes. A nested handler can
observe and mutate the actor's state mid-update, which relaxes the actor model's
guarantee of single-threaded sequential state mutation and admits the invariant
violations that guarantee is designed to prevent. The conservative policy avoids
this at the cost of rejecting some legitimate callback structures; the choice
between them is exposed as configuration rather than fixed.

\paragraph{Further evaluation.} The microbenchmarks of Section~\ref{sec:micro}
establish the per-operation latency reductions and the two-order-of-magnitude
tail benefit of the memory pool directly. The natural next step measures the
mechanism under load: synchronous versus asynchronous \emph{throughput} across
contention levels, chain depths, and message sizes on the x86-64 Linux target,
together with the distribution of synchronous completions versus asynchronous
fallbacks for \texttt{fast\_send\_void}. That saturated, multi-axis
characterization is the subject of ongoing work.

% ---------------------------------------------------------------
\section{Future work}
\label{sec:future}
% ---------------------------------------------------------------

\begin{enumerate}[leftmargin=1.6em]
  \item \textbf{Detecting cycles across threads} (Section~\ref{sec:cycle}),
  which a thread-local call chain cannot see: a global wait-for graph, a global
  lock order, or a bounded try-lock with asynchronous fallback.
  \item \textbf{The saturated-throughput and contention evaluation}
  (Section~\ref{sec:disc}).
  \item \textbf{Completing kaspar-hft:} the cyclic-invocation detector and its
  refinement, the three remaining delivery operations, the optimized
  synchronous-only reply, and enforcement of a single reply per message.
\end{enumerate}

% ---------------------------------------------------------------
\section{Conclusion}
% ---------------------------------------------------------------

The actor model has long been dismissed for high-frequency trading on a plausible
but mistaken premise: that its safety is inseparable from a discipline --- one
thread per actor, a mailbox per actor, a message and a context switch for every
interaction --- whose per-message cost cannot fit a microsecond budget. The
premise conflates the model with one way of implementing it. For \emph{co-located}
actors the discipline can be honored without that cost, and the resulting system
is fast enough for the HFT hot path while keeping the properties that motivate the
model: per-actor state isolation, freedom from data
races, and sequential single-message reasoning. The synchronous path gives up
part of the asynchronous model's freedom from deadlock: a cycle of synchronous
calls spread across threads deadlocks. This paper supports
that thesis both analytically and with a deployed, measured implementation ---
kaspar-hft --- rather than as theory alone.

Asynchronous message passing is intrinsic to the actor model's guarantees and
indispensable across machines, but for co-located actors it imposes heap
allocation, queue operations, scheduling, and context switches that isolation
does not require. \fs{} provides a synchronous alternative in which the sender's
thread runs the receiver's handler directly under the receiver's exclusive-access
lock and takes the reply as a value, while the handler's code does not depend on
the delivery mode, the executing thread, or the sender's blocked state ---
receiver transparency. A thread-local call-chain membership test, checked before
lock acquisition, catches synchronous cycles on one thread before they deadlock
or overflow the stack; cycles across threads need a different mechanism. A
refinement keyed on (actor, message-class) pairs admits legitimate callbacks
while still catching unbounded recursion, and five delivery operations expose the
latency/robustness spectrum behind one transparent handler interface. The performance case is made analytically, as the elimination of
per-message work, and corroborated by microbenchmarks in which the synchronous
path completes a round trip in tens of nanoseconds and a memory pool cuts the
latency tail by two orders of magnitude.

The applied study places these numbers against a live market-data path. On the
kaspar-hft framework, with the socket reader decoding each packet and updating
the book on its own thread, socket-to-book book-update medians are
$0.8$--$1.1~\mu$s, and the framework's per-hop cost ($\sim\!9$~ns, about one
unpredictable virtual call) is about $1\%$ of that. When the reader instead hands
each datagram to a second thread through an asynchronous mailbox, latency
decomposes into a $\sim\!7~\mu$s floor and a per-message serialization slope, the
floor stable to within a few tenths of a microsecond across three estimators and
a tenfold change in packet rate. The hand-off is the main structural difference
between the two paths, though the runs also differ in load and configuration, so
we do not attribute the whole difference to it. On the two-stage path part of
the tail is mailbox-queue backlog; the dominant effect is the market's
arrival process, which is non-Poisson and strongly clustered, and
converts bursts into large packets and in-packet
position into latency through the slope --- characterized in a companion
paper~\cite{mdarrival}. The parameters to optimize are therefore where the path
crosses threads and the per-message service time, not the message rate. The same study
also fixes the choice among the four mailbox queue implementations --- BQueue by
default, ShardedBQueue for high-fan-in actors --- and shows that a group of actors
sharing one queue is at once a real-time execution engine and a deterministic
simulator, so a trading strategy is developed in backtest and deployed to
production without modification.

Taken together, the mechanism and the measurement answer the question in the
title for the workload we tested. The four extensions --- \fs{}, actor groups,
per-actor mailbox selection, and the memory pool --- reduce the per-hop cost of
the actor framework on a live tick-to-book path to about one percent of the
$\sim\!1~\mu$s decode-and-book work when that path stays on one thread; an
asynchronous cross-thread hand-off in the path, by contrast, raises the floor and
adds a queueing tail. For this co-located market-data workload, kept on one
thread, the actor discipline therefore carries a small per-hop latency cost, and in return the design retains data-race freedom,
single-message reasoning (which we argue,
but do not here demonstrate, is amenable to automatic code generation), and a
production/simulation duality in which the same strategy code is backtested and
traded without change. These results suggest that actor-based designs need not be
excluded from HFT systems on the grounds of per-message messaging overhead ---
long the stated reason --- and that a deployed system runs on one. What this study
does not establish, and what remains for future work, is behavior under
saturation and contention: the evaluation here is sequential, single-machine, and
socket-timestamped, not a throughput or multi-core stress test. The remaining
work is listed in Section~\ref{sec:future}.

% ---------------------------------------------------------------


\begin{thebibliography}{99}
% ---------------------------------------------------------------
\bibitem{hewitt1973} C.~Hewitt, P.~Bishop, and R.~Steiger.
  \emph{A Universal Modular ACTOR Formalism for Artificial Intelligence.}
  Proc.\ 3rd International Joint Conference on Artificial Intelligence (IJCAI),
  1973.
\bibitem{agha1986} G.~Agha.
  \emph{Actors: A Model of Concurrent Computation in Distributed Systems.}
  MIT Press, 1986.
\bibitem{miller2005} M.~S.~Miller, E.~D.~Tribble, and J.~Shapiro.
  \emph{Concurrency Among Strangers: Programming in E as Plan Coordination.}
  Symposium on Trustworthy Global Computing, 2005.
\bibitem{ambienttalk2007} T.~Van Cutsem et al.
  \emph{AmbientTalk: Object-Oriented Event-Driven Programming in Mobile Ad Hoc
  Networks.} XXVI Int.\ Conf.\ of the Chilean Computer Science Society, 2007.
\bibitem{armstrong1996} J.~Armstrong, R.~Virding, C.~Wikstr\"om, and M.~Williams.
  \emph{Concurrent Programming in Erlang.} Prentice Hall, 2nd ed., 1996.
\bibitem{armstrong2003} J.~Armstrong.
  \emph{Making Reliable Distributed Systems in the Presence of Software Errors.}
  PhD thesis, Royal Institute of Technology (KTH), Stockholm, 2003.
\bibitem{akka} Lightbend, Inc.
  \emph{Akka Documentation.} \url{https://akka.io/docs/}.
\bibitem{caf2014} D.~Charousset, T.~C.~Schmidt, R.~Hiesgen, and M.~W\"ahlisch.
  \emph{CAF --- The C++ Actor Framework for Scalable and Resource-Efficient
  Applications.} Proc.\ 4th Int.\ Workshop on Programming based on Actors,
  Agents, and Decentralized Control (AGERE!), 2014.
\bibitem{orleans} S.~Bykov et al.
  \emph{Orleans: Cloud Computing for Everyone.}
  Proc.\ 2nd ACM Symposium on Cloud Computing (SoCC), 2011.
\bibitem{swift2021} Apple, Inc.
  \emph{Swift Evolution Proposal SE-0306: Actors.} 2021.
\bibitem{actor4j2018} D.~Bauer.
  \emph{Actor4j: A Software Framework for the Actor Model Focusing on the
  Optimization of Message Passing.} 2018.
\bibitem{kar2021} O.~Tardieu et al.
  \emph{Reliable Actors with Retry Orchestration.} IBM Research, 2021.
\bibitem{activeobject1996} R.~G.~Lavender and D.~C.~Schmidt.
  \emph{Active Object: An Object Behavioral Pattern for Concurrent Programming.}
  Pattern Languages of Program Design 2, Addison-Wesley, 1996.
\bibitem{posa2} D.~C.~Schmidt, M.~Stal, H.~Rohnert, and F.~Buschmann.
  \emph{Pattern-Oriented Software Architecture, Volume 2: Patterns for
  Concurrent and Networked Objects.} Wiley, 2000.
\bibitem{sutter2005} H.~Sutter and J.~Larus.
  \emph{Software and the Concurrency Revolution.} ACM Queue, 3(7):54--62, 2005.
\bibitem{linguafranca2023} M.~Lohstroh, C.~Menard, S.~Bateni, and E.~A.~Lee.
  \emph{High-Performance Deterministic Concurrency using Lingua Franca.}
  arXiv:2301.02444, 2023.
\bibitem{greif1975} I.~G.~Greif.
  \emph{Semantics of Communicating Parallel Processes.}
  PhD thesis, MIT, 1975. Project MAC Technical Report MAC-TR-154.
\bibitem{clinger1981} W.~D.~Clinger.
  \emph{Foundations of Actor Semantics.}
  PhD thesis, MIT, 1981. MIT AI Lab Technical Report AI-TR-633.
\bibitem{armstrong2007} J.~Armstrong.
  \emph{A History of Erlang.}
  Proc.\ 3rd ACM SIGPLAN Conference on the History of Programming Languages
  (HOPL III), 2007. \url{https://doi.org/10.1145/1238844.1238850}.
\bibitem{beambook} E.~Stenman.
  \emph{The Erlang Runtime System (The BEAM Book), chapter ``Scheduling''.}
  \url{https://github.com/happi/theBeamBook}, accessed 2026.
\bibitem{letitcrash} Lightbend (Akka Team).
  \emph{50 Million Messages per Second --- on a Single Machine.}
  \url{https://letitcrash.com/post/20397701710/50-million-messages-per-second-on-a-single},
  2012.
\bibitem{orleansmsr2014} S.~Bykov, A.~Geller, G.~Kliot, J.~R.~Larus, R.~Pandya,
  and J.~Thelin.
  \emph{Orleans: Distributed Virtual Actors for Programmability and
  Scalability.} Microsoft Research Technical Report MSR-TR-2014-41, 2014.
\bibitem{ponyorca2017} S.~Clebsch, J.~Franco, S.~Drossopoulou, A.~M.~Yang,
  T.~Wrigstad, and J.~Vitek.
  \emph{Orca: GC and Type System Co-design for Actor Languages.}
  Proc.\ ACM Program.\ Lang.\ 1(OOPSLA), Article~72, 2017.
\bibitem{camilleri2023} C.~Camilleri, J.~G.~Vella, and V.~Nezval.
  \emph{Actor Model Frameworks: An Empirical Performance Analysis.}
  In \emph{Advances in Information and Communication} (FICC 2023), Lecture Notes
  in Networks and Systems, vol.~652, Springer, 2023.
  \url{https://doi.org/10.1007/978-3-031-31153-6_37}.
\bibitem{imam2014savina} S.~M.~Imam and V.~Sarkar.
  \emph{Savina --- An Actor Benchmark Suite: Enabling Empirical Evaluation of
  Actor Libraries.} Proc.\ 4th Int.\ Workshop on Programming based on Actors,
  Agents, and Decentralized Control (AGERE!), 2014.
  \url{https://doi.org/10.1145/2687357.2687368}.
\bibitem{actix} Actix Project.
  \emph{Actix: Actor Framework for Rust --- Documentation.}
  \url{https://docs.rs/actix/}, accessed 2026.
\bibitem{tasharofi2013} S.~Tasharofi, P.~Dinges, and R.~E.~Johnson.
  \emph{Why Do Scala Developers Mix the Actor Model with Other Concurrency
  Models?} Proc.\ 27th European Conference on Object-Oriented Programming
  (ECOOP), LNCS 7920, Springer, 2013, pp.~302--326.
\bibitem{bagherzadeh2020} M.~Bagherzadeh, N.~Fireman, A.~Shawesh, and
  R.~Khatchadourian.
  \emph{Actor Concurrency Bugs: A Comprehensive Study on Symptoms, Root Causes,
  API Usages, and Differences.} Proc.\ ACM Program.\ Lang.\ 4(OOPSLA),
  Article~214, 2020. \url{https://doi.org/10.1145/3428282}.
\bibitem{torreslopez2018} C.~Torres Lopez, S.~Marr, H.~M\"ossenb\"ock, and
  E.~Gonzalez Boix.
  \emph{A Study of Concurrency Bugs and Advanced Development Support for
  Actor-based Programs.} In \emph{Programming with Actors}, LNCS 10789,
  Springer, 2018, pp.~155--185. Also arXiv:1706.07372.
\bibitem{concur2026} J.~Huang, T.~Mahmud, C.~P{\u a}s{\u a}reanu, and G.~Yang.
  \emph{CONCUR: Benchmarking LLMs for Concurrent Code Generation.}
  arXiv:2603.03683, 2026.
\bibitem{li2007} C.~Li, C.~Ding, and K.~Shen.
  \emph{Quantifying the Cost of Context Switch.} Proc.\ Workshop on Experimental
  Computer Science (ExpCS), ACM, 2007.
\bibitem{mcvoy1996} L.~McVoy and C.~Staelin.
  \emph{lmbench: Portable Tools for Performance Analysis.}
  Proc.\ USENIX Annual Technical Conference, 1996.
\bibitem{bendersky2018} E.~Bendersky.
  \emph{Measuring Context Switching and Memory Overheads for Linux Threads.}
  \url{https://eli.thegreenplace.net/2018/measuring-context-switching-and-memory-overheads-for-linux-threads/},
  2018.
\bibitem{kasparperf} V.~Maciejewski.
  \emph{kaspar-hft actor framework: C++ message-passing microbenchmarks
  (\texttt{actors/cpp/perf}).}
  \url{https://github.com/vincent212/kaspar-hft}, 2026.
\bibitem{fixsbe} FIX Trading Community.
  \emph{Simple Binary Encoding (SBE), Version 1.0.} Technical Standard, 2020.
  \url{https://www.fixtrading.org/standards/sbe/}.
\bibitem{shadowpov} V.~Maciejewski.
  \emph{Model-Free Passive Execution via Order-Level Shadowing.}
  arXiv:2609.18019, 2026. \url{https://arxiv.org/abs/2609.18019}.
\bibitem{foundationdb} J.~Zhou, M.~Xu, A.~Shraer, et al.
  \emph{FoundationDB: A Distributed Unbundled Transactional Key Value Store.}
  In Proc.\ ACM SIGMOD, 2021. \url{https://doi.org/10.1145/3448016.3457559}.
\bibitem{abides} D.~Byrd, M.~Hybinette, and T.~H.~Balch.
  \emph{ABIDES: Towards High-Fidelity Multi-Agent Market Simulation.}
  In Proc.\ ACM SIGSIM-PADS, 2020. \url{https://doi.org/10.1145/3384441.3395986}.
\bibitem{mdarrival} V.~Maciejewski.
  \emph{The Market-Data Arrival Process is Self-Exciting: Bursts, Packet
  Coalescing, and the Socket-to-Book Latency Tail.} kaspar-hft technical report
  (in preparation), 2026.
\end{thebibliography}
\end{document}